\documentclass[12pt]{article}
\makeatletter
\catcode`\@=11
\@addtoreset{equation}{section}

\makeatother

\baselineskip=\normalbaselineskip

\usepackage{mathrsfs,amsbsy,amssymb,latexsym,amsfonts,amsmath}
\usepackage{graphicx,color}
\usepackage{bm}
\usepackage{ulem}

\allowdisplaybreaks[4]

\title{Spin Dynamics with Inertia in Metallic Ferromagnets}
\author{Toru Kikuchi\footnote{toru.kikuchi@riken.jp}~ and Gen Tatara\\
\it{RIKEN Center for Emergent Matter Science (CEMS),}\\
 \it{2-1 Hirosawa, Wako, Saitama, Japan}}
\date{\today}

\begin{document}

  \maketitle

\begin{abstract}
Non-adiabatic contribution of environmental degrees of freedom yields effective inertia of spin in effective spin dynamics.  In this paper, we study several aspects of the inertia of spin in metallic ferromagnets. (i) a concrete expression of the spin inertia $m_s$: $m_s=\hbar S_c/(2g_{\rm sd})$, where $S_c$ is the spin polarization of conduction electrons and $g_{\rm sd}$ is the $sd$ coupling constant. (ii) dynamical behavior of spin with inertia, discussed from viewpoints of a spinning top and of a particle on a sphere. (iii) behavior of spin waves and domain walls in the presence of inertia, and behavior of spin with inertia in the case of a time-dependent magnetic field.  
\end{abstract}

\section{Introduction}

Different from other fundamental quantities such as mass and charge, spin is a dynamical quantity, 
and its dynamics have been widely studied and applied in science and technology.  In particular, recent rapid growth of spintronics provides a stage where deeper understandings of spin dynamics directly lead to practical applications. 

The dynamics of spin are governed by the spin Berry phase \cite{0}, and its equation of motion includes only the first-order time derivative of spin.  This is natural because spin is an angular momentum and its equation of motion takes the familiar form: the time-derivative of the angular momentum (i.e. spin) is given by the torque acting on it.   Without any torque, the solution of the equation of motion of spin is only a static one.  This is in contrast with, for example, the case of a massive point particle, which has inertia and can move at non-zero speed as its free motion.  In this sense, spin does not have inertia.   

However, for systems where spin interacts with other environmental degrees of freedom,  spin dynamics are affected by those environmental degrees of freedom, and the dynamical law of spin is changed to be an effective one.  For example, in metallic ferromagnets, conduction electrons affect the dynamics of localized spins (i.e. spins of atoms on lattice cites).   
A typical effect is 
spin-damping (e.g. \cite{Kohno2006}), where the energy and the angular momentum of spins are transferred to the environmental degrees of freedom and, as the result,  spins relax to their ground state within a certain time scale.   Without this effect, spins undergo the Larmor precession around the applied magnetic field forever.  In this respect, the existence of the degrees of freedom other than spins changes the dynamical behavior of spins significantly.   In the equation of motion of spins, this damping is represented by the Gilbert damping term \cite{Gilbert04}.  This damping term includes only the first-order time derivative of spins.  Therefore, spin still does not have inertia even when we take into account the Gilbert damping effect.    

The effects of the environments other than the Gilbert damping can be studied systematically by the derivative expansion, where the effects are expanded in powers of the time (and spatial) derivative of spin.  
 From that point of view, the Gilbert damping term gives the leading order term in that expansion.  In the higher orders, there appear terms which include the second-order time derivative, the third-order time derivative and so on, in the equation of motion of spin.  These terms with higher order time derivatives are interesting in that, like the Gilbert damping term, they change the dynamical law of spin itself, more than give additional torque on spin.  
 In particular, from a comparison with the form of the Newton's equation of motion of a massive point particle, the term with the second-order time derivative of spin plays the role of the inertia of spin.  

Such spin inertia has been discussed in the literature (e.g. ref.\cite{Suhl98}).  In particular, recent progress in ultrafast magnetization \cite{Beaurepaire96, Kirilyuk10} motivated several works.  In refs.\cite{Ciornei11, Olive12}, the inertia of spin was introduced phenomenologically and was shown to give additional nutation to the motion of spin.   The time scale where the effect of the inertia is significant was discussed, based on the work of Brown \cite{Brown63}, to be sub-picosecond order.    In ref.\cite{Wegrowe12}, the equivalence between the dynamics of spin with inertia and a spinning top was discussed.   Microscopic derivation of spin inertia was performed in refs.\cite{Fahnle11} and \cite{Bhattacharjee12}.  In ref.\cite{Fahnle11}, an extended breathing Fermi surface model was used, and the relation between the Gilbert damping coefficient and the spin inertia was given in terms of physical quantities (such as Fermi-Dirac occupation numbers) of conduction electrons.  The time scale for the nutational motion to be damped by the Gilbert damping was estimated to be sub-picosecond order.  In ref.\cite{Bhattacharjee12}, a general expression of the contribution of the conduction electrons to spin dynamics was discussed.  The spin effective dynamics was shown to be non-local in general, which can be approximated as local dynamics by the derivative expansion of spin.  The inertial term of spin arises in that derivative expansion and its general expression was given in term of the Green's function of the conduction electrons.  

Since the inertia of spin is conceptually interesting in its own and gives the first step toward the understanding of non-adiabatic contribution of environmental degrees of freedom to spin effective dynamics, further investigations are worthwhile.   Although the expressions of the inertia of spin were given as integral forms \cite{Fahnle11, Bhattacharjee12}, an explicit expression of the inertia of spin in terms of parameters of a model has not been obtained so far.  Furthermore, the effect of the inertia of spin on the dynamics of spin has been discussed only for spatially homogeneous spin system under time-independent magnetic field.  
In this paper, we present detailed theoretical study of the effects induced by the spin inertia based on an $sd$ model.  
In section \ref{sec: effective action}, we derive a concrete expression of spin inertia in terms of the parameters in the $sd$ model.  In section \ref{sec: spin and spinning top}, the basic behavior of spin with finite inertia is studied with the help of its two equivalents: a symmetric spinning top and a massive charged particle on a sphere subject to a monopole field.  
In section \ref{sec: phenomena}, we study spatially inhomogeneous system, and discuss that spin waves and magnetic domain walls acquire an additional oscillation mode due to spin inertia.  We also study the behavior of spin under large and time-dependent magnetic field, and find an unusual behavior of spin where the velocity of spin is parallel to the direction of the time-derivative of magnetic field. 

\section{Spin effective action and inertia}\label{sec: effective action}

The inertia of spin arises naturally in its effective dynamics, which takes into account the effects of the environmental degrees of freedom \cite{Bhattacharjee12}.  Consider a system where a classical field of localized spin, $\bm S(\bm x,t)=S_l \bm n(\bm x,t)$ (with $S_l=|\bm S|$ fixed), and a field $c(\bm x,t)$ representing the other environmental degrees of freedom are interacting with each other.  For concreteness, we consider the case of metallic ferromagnets in this paper.  Conduction electrons are represented by annihilation and creation operators, $c$ and $\bar c$.  The total action is given by $\mathcal S_{\rm s}[\bm n]+\mathcal S_{\rm e}[c, \bar c, \bm n]$ where $\mathcal S_{\rm s}[\bm n]$ is the action of spin $\bm n$ and $\mathcal S_{\rm e}[c, \bar c, \bm n]$ is that of conduction electrons with their interaction with spin $\bm n$.  When we are interested only in the dynamics of spin $\bm n$, it is convenient to integrate out $c$ and $\bar c$, and derive the effective action of spin $\bm n$.  The contribution from electrons is given by path integration as
\begin{equation}
 \exp\left(\frac{i}{\hbar}\Delta \mathcal S_{\rm eff}[\bm n]\right) 
\equiv   \int \mathcal D \bar c\mathcal D  c ~ \exp\left(\frac{i}{\hbar}\mathcal S_{\rm e}[c, \bar c, \bm n]\right).
\end{equation}
The sum of this $\Delta \mathcal S_{\rm eff}[\bm n]$ and the original spin action $\mathcal S_{\rm s}[\bm n]$ gives the total spin effective action.  

It is difficult to calculate $\Delta \mathcal S_{\rm eff}[\bm n]$ exactly, so we should rely on a perturbative analysis.  We here perform derivative expansion, where $\Delta \mathcal S_{\rm eff}[\bm n]$ is expanded in powers of $\partial_\mu \bm n$ ($\mu=t,x,y,z$).  When the system is isotropic, the general form is  
\begin{equation}
\Delta \mathcal S_{\rm eff}[\bm n] = \int \frac{d^3x}{a^3} dt\left[ S_c\dot\phi(\cos\theta-1)  - \frac{J_cS_c^2}{2}(\partial_i \bm n)^2 + \frac{m_s}{2}\bm {\dot n}^2 \right] + \mathcal O((\partial_\mu \bm n)^3),
\label{Delta Seff in main text}
\end{equation}
where $\bm n=(\sin\theta\cos\phi,\sin\theta\sin\phi,\cos\theta)$ and $i=x,y,z$.  We have divided the Lagrangian density entirely by lattice volume $a^3$ so that each coefficient represents a quantity per each lattice cite.   The first term is the spin Berry phase with $S_c$ the spin polarization of the conduction electrons and the second is the spin-spin exchange interaction induced by electrons with $J_c$ the coupling constant.  In the final term, there arises the inertial term of spin with the inertia $m_s$.  This $m_s$ has the dimension of $[{\rm kg}\cdot {\rm m}^2]$, the same as that of the moment of inertia.  
In ref.\cite{Bhattacharjee12}, a general expression of spin inertia $m_s$ is derived in term of the Green's function of the conduction electrons.  Let us here calculate a concrete expression of spin inertia.  
As in ref.\cite{Bhattacharjee12}, a typical example of $\mathcal S_{\rm e}[c,\bar c, \bm n]$ is the $sd$ model \cite{Tatara_reports}, where conduction electrons interact with localized spins $\bm n$ as
\begin{equation}
\mathcal S_{\rm e}[c, \bar c, \bm n] = \int d^3x dt~ \bar c
\left(i\hbar \partial_t + \frac{\hbar^2\partial_i^2}{2m} + \epsilon_F + g_{\rm sd} \bm n\cdot \bm \sigma\right)c.
\label{s-d model}
\end{equation}
Here, $m$ is the mass of the conduction electrons, $\epsilon_F$ is the Fermi energy, $g_{\rm sd}$ is the $sd$ coupling constant, and $\bm \sigma$ is the Pauli matrix vector.  To obtain the derivative expansion of $\Delta \mathcal S_{\rm eff}[\bm n]$, we perform $SU(2)$ gauge transformation $c\rightarrow U(\bm x,t)c$ with an $SU(2)$ matrix $U$ acting on the spinor indices, so that the $sd$ interaction becomes diagonal, $\bar c (\bm n\cdot \bm \sigma)c \rightarrow \bar c \sigma_3 c$.  Due to this unitary transformation, there appears a so-called spin gauge field $A_\mu \equiv -iU^\dagger\partial_\mu U$ in \eqref{s-d model} through $\partial_\mu c \rightarrow U(\partial_\mu + iA_\mu)c$.  This $A_\mu$ contains the first-order derivative $\partial_\mu \bm n$.  Therefore, expanding $\exp(i\mathcal S_{\rm e}[c, \bar c, \bm n]/\hbar)$ in powers of $A_\mu$, we can calculate $\Delta \mathcal S_{\rm eff}[\bm n]$ perturbatively in powers of $\partial_\mu \bm n$.  The spin polarization $S_c$ and the inertia $m_s$ can be  calculated as (see Appendix \ref{sec: calculation of spin effective action} for details)
\begin{equation}
S_c = a^3 \frac{\hbar}{2}\frac{k_{F+}^3-k_{F-}^3}{6\pi^2},~~~m_s = \frac{\hbar S_c}{2g_{\rm sd}} 
\label{expression of inertia}
\end{equation}
with $\hbar^2k_{F\pm}^2/(2m) \equiv \epsilon_F \pm g_{\rm sd}$.  
The inertia can be rewritten as 
\begin{equation}
m_s = (k_F a)^3 \frac{\hbar^2}{8\pi^2}\frac{1}{\epsilon_F}f(g_{\rm sd}/\epsilon_F),~~ 
f(x)\equiv \frac{1}{3x}\left[(1+x)^{\frac{3}{2}}-(1-x)^{\frac{3}{2}}\right].
\label{expression of inertia 2}
\end{equation}
For $0<x<1$, the function $f(x)$ is a only slightly decreasing function from $f(0)=1$ to $f(1)\fallingdotseq 0.94$. Therefore, spin inertia does not depend much on the $sd$ coupling constant $g_{\rm sd}$ and is proportional to $a^3\sqrt{\epsilon_F}$, when $a$ and $k_F=\sqrt{2m\epsilon_F}/\hbar$ are regarded as independent parameters.

Adding this $\Delta \mathcal S_{\rm eff}[\bm n]$ to the original spin action of the form 
\begin{equation}
\mathcal S_{\rm s}[\bm n] = \int \frac{d^3x}{a^3} dt\left[ S_l\dot\phi(\cos\theta-1) - \frac{J_lS_l^2}{2}(\partial_i \bm n)^2  \right],
\label{spin original action}
\end{equation}
with $J_l$ the exchange coupling between spins, we obtain the total spin effective action as $\mathcal S_{\rm eff}[\bm n] = \mathcal S_{\rm s}[\bm n]+\Delta \mathcal S_{\rm eff}[\bm n]$.  Including the Zeeman coupling with external magnetic field\footnote{To be precise, there emerge other coupling terms between spin and electromagnetic field in addition to the Zeeman coupling, such as in ref.\cite{Kawaguchi14}, by integrating out conduction electrons.  In this paper, we simply assume that they are negligible. }
we obtain 
\begin{equation}
\mathcal S_{\rm eff}[\bm n] = \int \frac{d^3x}{a^3} dt\left[S\bm B \cdot \bm n + S\dot\phi(\cos\theta-1)  - \frac{JS^2}{2}(\partial_i \bm n)^2 + \frac{m_s}{2}\bm {\dot n}^2 \right] + \mathcal O((\partial_\mu \bm n)^3),
\label{spin effective action}
\end{equation}
with $S\equiv S_c+S_l$ the total spin amplitude per lattice cite and $J\equiv (J_cS_c^2+J_lS_l^2)/S^2$.  We have set the gyromagnetic ratio as unity.  

As we will see below, spin with finite inertia has a typical precession mode with frequency $\omega_0\sim S/m_s$.  Using $m_s$ \eqref{expression of inertia} or \eqref{expression of inertia 2}, and assuming $S\sim \hbar$, $k_F a \sim \pi$ and $\epsilon_F\sim 1{\rm eV}$, the energy scale of this frequency becomes $\hbar \omega_0 \sim 1{\rm eV}$, so that its period is $2\pi/\omega_0\sim 0.1{\rm ps}$.  Therefore, as far as this simple estimation suggests, the existence of the inertia is significant for the dynamics of sub-picosecond scale.  

The equation of motion derived from this effective action \eqref{spin effective action} is 
\begin{equation}
S\bm{\dot n} = -S\bm B\times \bm n  - JS^2\partial_i^2 \bm n\times  \bm n+ m_s\bm{\ddot n} \times \bm n. 
 \label{spin EOM}
\end{equation}
Thus, the inertial term produces acceleration-dependent torque.  We can rewrite this equation of motion, by taking vector product with $\bm n$, as 
\begin{equation}
m_s \bm{\ddot n} = S\bm n \times \bm{\dot n} + S\bm B + JS^2\partial_i^2\bm n
 - (S\bm B\cdot \bm n + m_s \bm{\dot n}^2 - JS^2(\partial_i \bm n)^2)\bm n.
\label{real spin EOM}
\end{equation}
The Gilbert damping effect adds a term $-\alpha S\bm {\dot n}$, with $\alpha$ the dimensionless constant, to the right hand side of the equation of motion \eqref{real spin EOM}.   Therefore, the Gilbert damping plays the same role as the familiar linear damping force for a point particle, and  the time scale for this damping term to be significant is $t_{\rm damp}\sim m_s/(\alpha S)$.  On the other hand, the time scale for the inertial term to be effective is, as we will see below, $t_{\rm inertia}\sim m_s/S$.  Therefore, for $\alpha\ll 1$, we can neglect the Gilbert damping term as long as we are interested in the dynamics within the time scale $t_{\rm inertia}$.    

The equation of motion \eqref{spin EOM} can be rewritten (when $\bm B=0$) in the conservation form of the angular momentum current $(\bm j^0,\bm j^i)$, 
\begin{equation}
\partial_0 \bm j^0 + \partial_i \bm j^i = 0, ~~~{\rm where}~
\bm j^0 = S\bm n + m_s \bm n \times \bm {\dot n},~ \bm j^i = JS^2 \partial_i \bm n \times \bm n
\label{angular momentum conservation}
\end{equation}
($\partial_0\equiv \partial/\partial t$).  Note that the angular momentum $\bm j^0$ (per lattice cite), which is the Noether charge corresponding to the invariance of the action \eqref{spin effective action} under $SO(3)$ rotation in the internal spin space (see Appendix \ref{sec: derivation of the angular momentum} for details), is no longer proportional to $\bm n$ but includes the non-adiabatic contribution of the conduction electrons, $m_s \bm n\times \bm {\dot n}$.  
Originally, the total angular momentum consists of that of the localized spin and that of the conduction electrons: $\bm j^0 = S_l \bm n + (\hbar a^3/2)\langle \bar c \bm \sigma c \rangle$ with $S_l$ the amplitude of the localized spin.  In the lowest order of the derivative expansion, i.e. in the adiabatic limit, the spin of the conduction electron aligns with that of the localized spin, so that $(\hbar a^3/2)\langle \bar c \bm \sigma c \rangle = S_c \bm n$ with $S_c$ the spin polarization of the conduction electrons.  Beyond the adiabatic limit, the direction of the spin of the conduction electron is generally different from that of the localized spin.  The derivative expansion incorporates this difference systematically, and the next order term in $(\hbar a^3/2)\langle \bar c \bm \sigma c \rangle$ is given by $m_s\bm n\times \bm{\dot n}$.

Remarkably, the relation between the inertia $m_s$ and the spin polarization $S_c$ in \eqref{expression of inertia}, $m_s=\hbar S_c/(2g_{\rm sd})$, is easily obtained without any microscopic calculation, as follows.   As we have discussed in the last paragraph, the angular momentum $\Delta \bm j^0$ derived from $\Delta \mathcal S_{\rm eff}[\bm n]$ represents the spin polarization of the conduction electrons, 
\begin{equation}
\Delta \bm j^0\equiv S_c \bm n + m_s \bm n \times \bm {\dot n}=\frac{\hbar a^3}{2}\langle \bar c \bm \sigma c \rangle.
\label{delta j}
\end{equation}
Since $(\hbar a^3/2)\langle \bar c \bm \sigma c \rangle$ obeys the following  equation of motion,
\begin{equation}
\partial_0\left(\frac{\hbar a^3}{2}\langle \bar c \bm \sigma c \rangle\right) = -\frac{2g_{\rm sd}}{\hbar}\bm n\times \left(\frac{\hbar a^3}{2}\langle \bar c \bm \sigma c \rangle \right)
\label{equation for c}
\end{equation}
(we consider here only spatially homogeneous case, for simplicity), $\Delta \bm j^0$ also satisfies 
\begin{equation}
\partial_0\Delta \bm j^0 = -\frac{2g_{\rm sd}}{\hbar}\bm n\times \Delta \bm j^0.
\label{equation of delta j}
\end{equation}
Substitution of eq.\eqref{delta j} into eq.\eqref{equation of delta j} leads to
\begin{equation}
\left(S_c-\frac{2g_{\rm sd}}{\hbar}m_s\right)\bm {\dot n} + \mathcal O((\partial_0)^2) = 0.
\end{equation}
Since this equation is true for an arbitrary $\bm n$, we arrive at the relation $m_s=\hbar S_c/(2g_{\rm sd})$.  Thus, we can obtain $m_s$ from $S_c$ without any detailed calculation.  The point is that the equation \eqref{equation of delta j} is not the equation of motion of $\bm n$, although it involves the time derivative of $\bm n$. [The equation of motion of $\bm n$ is $\partial_0\bm j^0=0$, or, $\partial_0(S_l\bm n + \Delta \bm j^0)=0$.]  Before integrating out the electrons, it was the equation of motion of $\langle \bar c \bm \sigma c \rangle$ \eqref{equation for c} and, after integrating out the electrons, the  equation \eqref{equation of delta j} determines the structure of $\Delta \mathcal S_{\rm eff}[\bm n]$.  
Conversely, the relation $m_s=\hbar S_c/(2g_{\rm sd})$ must hold in order that the effective action $\Delta\mathcal S_{\rm eff}$ reproduces the equation of motion of $\langle \bar c \bm \sigma c \rangle$.  We can repeat this procedure to arbitrary orders in the derivative expansion:  first, write down all possible terms in the effective action $\Delta \mathcal S_{\rm eff}$, with their coefficients left undetermined;  second, derive the angular momentum $\Delta \bm j^0$ from that $\Delta \mathcal S_{\rm eff}$ and substitute it into eq.\eqref{equation of delta j};  then we can obtain the recursion relations between the coefficients\footnote{Such recursion relations can be obtained also via the equation of motion of spin, as follows.  The original equation of motion of localized spin is $S_l \bm{\dot n} = -S_l\bm B\times \bm n - J_lS_l^2\partial_i^2\bm n\times \bm n -g_{\rm sd}a^3\langle \bar c \bm \sigma c\rangle \times \bm n$.   Substitution of the expression of $\langle \bar c \bm \sigma c\rangle$ [eq.\eqref{delta j}] into this original equation of motion gives the term proportional to $\bm{\dot n}$, i.e. $-g_{\rm sd}a^3\langle \bar c \bm \sigma c\rangle\times \bm n=-(2g_{\rm sd}m_s/\hbar)\bm{\dot n}+\mathcal O(\partial_0^2)$. The coefficient of this term is identical with the spin polarization $S_c$ of the conduction electron, which gives the renormalization of the spin amplitude, $S_l\rightarrow S_l+S_c$.  Therefore, we obtain the relation $m_s=\hbar S_c/(2g_{\rm sd})$.  }
.  For example, $\Delta \mathcal S_{\rm eff}$ to the fourth-order can be obtained as follows (see Appendix \ref{sec: effective action to the fourth} for details):
\begin{align}
\Delta \mathcal S_{\rm eff}[\bm n]=&S_c\int \frac{d^3x}{a^3}dt\Big[
\dot\phi(\cos\theta-1) 
+ \frac{1}{4}\frac{\hbar}{g_{\rm sd}}\bm{\dot n}^2 \notag \\
&~~~~~~- \frac{1}{8}\left(\frac{\hbar}{g_{\rm sd}}\right)^2\bm n\cdot(\bm{\dot n}\times\bm{\ddot n}) 
+ \frac{1}{16}\left(\frac{\hbar}{g_{\rm sd}}\right)^3\bm{\ddot n}^2 
- \frac{5}{64}\left(\frac{\hbar}{g_{\rm sd}}\right)^3(\bm{\dot n}^2)^2
\Big] + \mathcal O(\partial_0^5).
\label{Delta Seff to the fourth}
\end{align}   
The same procedure can be applied also for the terms in the effective action $\Delta \mathcal S_{\rm eff}$ which involve the spatial derivative.  

\section{Dynamical behavior of spin with inertia}\label{sec: spin and spinning top}

\begin{figure}[tbp]
\begin{center}
\includegraphics[width=150mm]{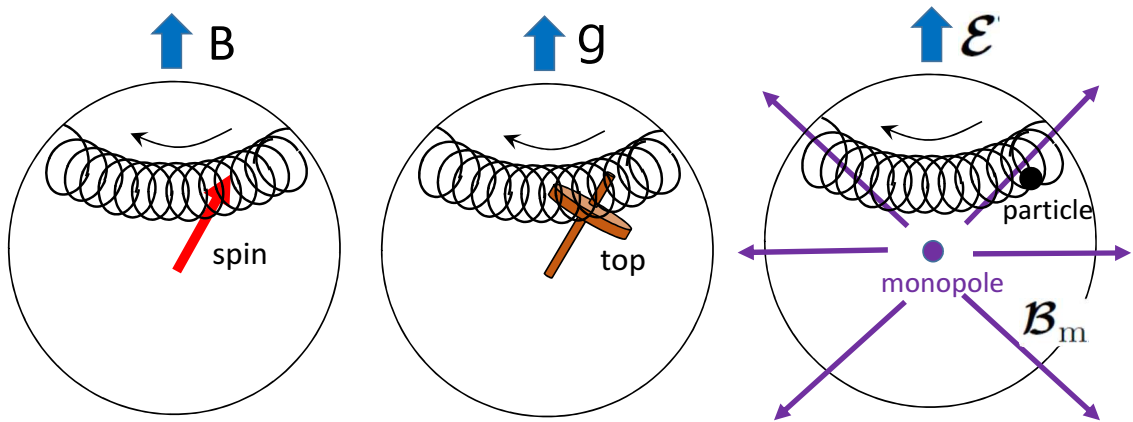}
\caption{\footnotesize{The dynamics of spin with inertia, a symmetric spinning top, and a massive charged particle on a sphere subject to a monopole magnetic field $\bm{\mathcal B}_{\rm m}$, are classically equivalent.  They undergo precession motion accompanied by nutation under applied fields, which are a magnetic field $B$ for spin, a gravitational field $g$ for the top, and an electric field $\mathcal E$ for the particle, respectively.  See the main text for the detail. }  }
\label{fig: spin top particle}
\end{center}
\end{figure}

In this section, we describe classical dynamics of spin with inertia.  For that purpose, it is helpful to use two equivalent pictures, which are summarized in Fig.\ref{fig: spin top particle}.  One is a symmetric spinning top, and the other is a massive charged particle on a sphere subject to a monopole magnetic field.  We consider here only spatially homogeneous spin, $\partial_i \bm n=0$, under time-independent magnetic field, for simplicity.     

\subsection{Equivalence to a symmetric spinning top}

The equivalence between the classical dynamics of spin and a spinning top has been recognized in the literature, e.g. refs.\cite{Morrish01} and \cite{Wegrowe12}.  
The content of this subsection is essentially a recapitulation of these facts, which we describe here in order for this paper to be self-contained.  

First, let us write down the Lagrangian of spin with finite inertia and that of a spinning top.  The Lagrangian of spin \eqref{spin effective action} is
\begin{equation}
L_{\rm spin}=\frac{m_s}{2}(\dot \theta^2 + \dot \phi^2\sin^2\theta ) + S\dot \phi (\cos\theta-1) + B S \cos\theta,
\label{L of spin}
\end{equation}
while the Lagrangian of a spinning top in terms of the Euler angles ($\theta, \phi, \psi$) is 
\begin{equation}
L_{\rm top}=\frac{I_1}{2}(\dot \theta^2 + \dot \phi^2\sin^2\theta ) 
	+ \frac{I_3}{2}(\dot\psi + \dot \phi \cos\theta )^2 + \mu g l \cos\theta,
\label{L of top}
\end{equation}
where $I_1, I_3$ are the principle moments of inertia, $\mu$ is the mass of the top, $g$ is the  gravitational acceleration constant, and $l$ is the distance between the center of mass and the fixed extremity of the top.  Here we have taken a symmetric spinning top and set two moments of inertia equal, $I_1=I_2$.  We take the positive directions of the external magnetic field $\bm B$ and the gravity both in the positive $z$ direction.  Note that $L_{\rm spin}$ is a function of $(\theta,\phi)$ while $L_{\rm top}$ is that of $(\theta,\phi,\psi)$.  Let us see below that they have equivalent dynamics concerning $(\theta,\phi)$.  

The equivalence can be directly seen at the level of their equations of motion.  The equation of motion of spin is 
\begin{align}
& m_s (\ddot \theta -\dot \phi^2\sin\theta\cos\theta) + S\dot \phi \sin\theta + B S\sin\theta = 0,
\notag \\
& \frac{d}{dt}\left[ m_s \dot \phi \sin^2\theta + S(\cos\theta-1) \right] = 0,
\label{eom of quadratic spin}
\end{align}
while the equation of motion of the spinning top is
\begin{align}
& I_1 (\ddot \theta -\dot \phi^2\sin\theta\cos\theta) 
+ I_3(\dot\psi+\dot\phi\cos\theta)\dot\phi\sin\theta \notag + \mu g l \sin\theta = 0, \notag \\
& \frac{d}{dt}\left[ I_1\dot\phi\sin^2\theta +I_3 (\dot\psi + \dot\phi\cos\theta)\cos\theta  \right] = 0, \notag \\
& \frac{d}{dt}\left[ I_3 (\dot\psi+\dot\phi\cos\theta) \right] = 0.
\label{eom of spinning top}
\end{align}
From the last equation in eq.\eqref{eom of spinning top}, the canonical momentum $M_3$ conjugate to $\psi$,
\begin{equation}
M_3\equiv I_3 (\dot\psi+\dot\phi\cos\theta),
\label{M3}
\end{equation}
is conserved.  Substituting this $M_3$ for $\dot \psi$ in the other two equations in \eqref{eom of spinning top}, we obtain the same equations as \eqref{eom of quadratic spin} with replacements
\begin{equation}
m_s \leftrightarrow I_1, ~~S \leftrightarrow M_3, ~~B S \leftrightarrow  \mu g l.
\label{replacements}
\end{equation}
Thus, the classical behaviors of $\theta$ and $\phi$ are the same for spin and a spinning top.   

We can see the correspondence more explicitly through their Hamiltonians :  
\begin{equation}
H_{\rm spin} = \frac{p_\theta^2}{2m_s} 
+ \frac{1}{2m_s}\frac{(M_\phi-S\cos\theta)^2}{\sin^2\theta} - BS\cos\theta
\label{H of spin}
\end{equation}
and 
\begin{equation}
H_{\rm top} = \frac{p_\theta^2}{2I_1} 
+ \frac{1}{2I_1}\frac{(M_\phi-M_3\cos\theta)^2}{\sin^2\theta} - \mu g l\cos\theta 
+ \frac{M_3^2}{2I_3},
\label{H of top}
\end{equation}
where $p_\theta \equiv \partial L / \partial \dot \theta$ and $M_\phi \equiv \partial L / \partial \dot \phi$ are the canonical momenta of $\theta$ and $\phi$, respectively.  These two Hamiltonians are completely the same under the replacements \eqref{replacements}.  
[The last term in \eqref{H of top} does not contribute to the dynamics of $\theta$ and $\phi$.]

The Lagrangians \eqref{L of spin} and \eqref{L of top} are related by the Legendre transformation about $\psi$:
\begin{equation}
L_{\rm spin}(\theta,\dot\theta,\dot\phi~;S) = L_{\rm top}(\theta,\dot\theta,\dot\phi,\dot\psi) - S\dot \psi~|_{\dot\psi=\dot\psi(\theta,\dot\phi,S)}
\label{Legendre}
\end{equation}
with replacements \eqref{replacements}.  In the right hand side, $\dot\psi$ is substituted by $S$ (or $M_3$) via \eqref{M3}.  The situation is quite similar to that of the familiar centrifugal force problem.  There, the original Lagrangian is given as $L(r,\dot r,\dot\phi)=(m/2)(\dot r^2 + r^2\dot \phi^2)-U(r)$, and we can obtain $\phi$-reduced Lagrangian by the Legendre transformation about $\phi$: $L_{\rm red}(r,\dot r; M)\equiv L-M\dot\phi = (m/2)\dot r^2 - M^2/(2mr^2)-U(r)$.  In exchange for reducing $\phi$, there appears a fictitious potential $M^2/(2mr^2)$.  Likewise, spin dynamics is the $\psi$-reduced dynamics of a spinning top, and the spin Berry phase (the second term in eq.\eqref{L of spin}) appears as the fictitious potential arising from the $\psi$-reduction.  When we perform the Legendre transformation also about $\theta$ and $\phi$ on both sides in eq.\eqref{Legendre}, we are led to the same Hamiltonians \eqref{H of spin} and \eqref{H of top}.

\begin{figure}[tbp]
\begin{center}
\includegraphics[width=120mm]{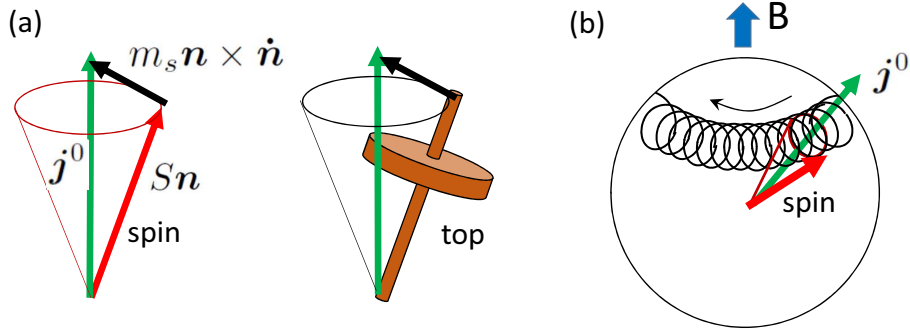}
\caption{\footnotesize{(a) Without any magnetic field, general motion of spin with finite inertia is a free precession motion, just like a spinning top.  The spin $S\bm n$ precesses around the total angular momentum $\bm j^0$.  (b) When a constant magnetic field $\bm B$ is applied, the total angular momentum $\bm j^0$ precesses around $\bm B$.  Therefore, the \textquoteleft free precession cone' in (a) precesses around $\bm B$ as a whole, which corresponds to the Larmor precession in the absence of the inertia.  What was called the free precession  in (a) is now called the nutation. }  }
\label{fig: free precession and nutation}
\end{center}
\end{figure}

Since the classical dynamics of spin with inertia and of a spinning top are equivalent, spin with inertia behaves in the same manner as a spinning top does (Fig.\ref{fig: free precession and nutation}).  When a magnetic field is not applied, spin undergoes free precession: the spin precesses around the total angular momentum $\bm j^0=S\bm n + m_s\bm n \times \bm{\dot n}$ (eq.\eqref{angular momentum conservation}), which is a constant of motion.  When a magnetic field is applied, spin undergoes the Larmor precession accompanied by the nutation:  the total angular momentum $\bm j^0$  precesses around the magnetic field (the Larmor precession), and the spin precesses around $\bm j^0$ at each time (the nutation) \cite{Ciornei11}. 
The free precession solution of the equation of motion \eqref{eom of quadratic spin} (with $\bm B=0$) is
\begin{equation}
\theta=\theta_0~ ({\rm const.}), ~~\dot \phi= \frac{S}{m_s \cos\theta_0}.
\label{free precession solution}
\end{equation}
Therefore, the free precession frequency, or the nutation frequency, $\omega_0$, is $\omega_0\sim S/m_s$, assuming that $\cos\theta_0$ is order of unity (that is, the radius of the spin free precession or nutation is not very large).  This gives the typical time scale for the inertial term, $t_{\rm inertia}\sim m_s/S$.    

Finally, we mention that usual spin with zero inertia, $m_s = 0$, can be regarded as follows.  It corresponds to the case of $I_1(=I_2)=0$ under replacements \eqref{replacements}.  This means that the other principle moment of inertia, $I_3$, vanishes since, for an actual rigid body, any one of the principle moments of inertia is equal to or less than the sum of the other two, e.g., $I_3 \leq I_1 + I_2$ \cite{Landau-Lifshitz}.  Thus, setting $I_1=I_2=0$ leads to $I_3=0$.  Therefore, 
there does not exist a spinning top corresponding to spin with zero inertia, in a non-relativistic framework.  

\subsection{Equivalence to a massive charged particle}

Since the spin Berry phase corresponds to a monopole gauge field \cite{0},   the classical dynamics of spin
is also equivalent to that of a charged particle on a sphere subject to a monopole background.  
Let us describe this equivalence and use it to understand the behavior of spin with inertia.  

A magnetic monopole yields a magnetic field in the radial direction with the strength inversely proportional to the square of the distance from the monopole.  Being put at the origin $\bm x=0$, the gauge field $\bm {\mathcal A}_{\rm m}$ and the magnetic field $\bm {\mathcal B}_{\rm m}$ of the monopole are 
\begin{equation}
\bm {\mathcal A}_{\rm m} = q\frac{1-\cos \theta}{r\sin\theta}\bm e_\phi,~~~\bm {\mathcal B}_{\rm m}= \bm \nabla\times \bm {\mathcal A}_{\rm m} =  \frac{q}{r^3}\bm x,
\end{equation}
where $\bm x=r(\sin\theta\cos\phi,\sin\theta\sin\phi,\cos\theta)$, $\bm e_\phi$ is the unit vector in $\phi$-direction, and $q$ is the magnetic charge of the monopole.  The Lagrangian of a massive charged particle coupled to this monopole magnetic field and a constant electric field $\bm {\mathcal E}$ is 
\begin{align} 
L =& \frac{m}{2}\bm {\dot x}^2 - \bm {\mathcal A}_{\rm m}\cdot \bm{\dot x} -  \Phi
\notag \\
=& \frac{m}{2}\bm {\dot x}^2 + q\dot \phi (\cos\theta-1) + \bm {\mathcal E}\cdot \bm x,
\end{align}
where we take the electromagnetic scalar potential $\Phi=-\bm{\mathcal E}\cdot \bm x$ so that $-\bm \nabla \Phi = \bm {\mathcal E}$.  This Lagrangian is identical with that of spin with inertia \eqref{spin effective action} when we assume that the particle is constrained on a unit sphere and identify the direction of the particle $\bm x$ with the spin direction $\bm n$, the mass $m$ with the spin inertia $m_s$, the magnetic charge $q$ with the spin amplitude $S$, and the electric field $\bm {\mathcal E}$ on the particle with the magnetic field $S\bm B$ on spin (we use calligraphic letters for electromagnetic fields on the particle).  

Being restricted on the unit sphere, the dynamical degrees of freedom of the particle are the spherical angles $\theta$ and $\phi$, and the equation of motion is, in vectorial form,  
\begin{align}
&\bm e_\theta \left(\frac{d}{dt}\frac{\partial L}{\partial \dot \theta}-\frac{\partial L}{\partial \theta}\right) +  \frac{\bm e_\phi}{\sin\theta} \left(\frac{d}{dt}\frac{\partial L}{\partial \dot \phi}-\frac{\partial L}{\partial \phi}\right) = 0 \notag \\
{\rm or}~~~~&
m \bm{\ddot x} = q\bm x \times \bm{\dot x} + \bm {\mathcal E}
 - (\bm {\mathcal E}\cdot \bm x + m \bm{\dot x}^2 )\bm x,
\end{align}
which is identical with \eqref{real spin EOM} (with $\partial_i\bm n=0$).  The last term proportional to $\bm x$ in the right hand side of the second line is the constraint force,  which keeps $|\bm x|=1$.
The particle is on a unit sphere and is subject to the monopole magnetic field $\bm {\mathcal B}_{\rm m}=q\bm x$ (on the sphere $|\bm x|=1$) and the electric field $\bm {\mathcal E}$.  The general motion of the particle is, in the absence of the electric field $\bm {\mathcal E}$, the cyclotron motion, and, in the presence of the electric field $\bm {\mathcal E}$, the $\bm {\mathcal E}\times \bm {\mathcal B}_{\rm m}$ drift motion\footnote{Our convention for the sign of magnetic field is such that the equation of motion of a particle is $m\bm{\ddot x}=\bm {\mathcal B}\times \bm {\dot x} + \bm {\mathcal E}$, which yields drift motion of the particle in $\bm {\mathcal B}\times \bm{\mathcal E}$ direction, rather than $\bm {\mathcal E}\times \bm{\mathcal B}$ direction.}.  

In view of the equivalence between the dynamics of spin with inertia and of the charged particle on a sphere, the free precession of spin described in the last subsection corresponds to the cyclotron motion of the particle with the frequency $\omega_0\sim q/m = S/m_s$, and the Larmor precession of spin accompanied by nutation corresponds to the $\bm {\mathcal E}\times \bm {\mathcal B}_{\rm m}$ drift motion.  

We summarize here the three pictures for the classical dynamics described by \eqref{spin effective action}.  
 
\begin{tabular}{c| c c}
\hline 
picture & inertia & intrinsic magnetic field \\
\hline\hline
spin & the inertia $m_s$  & the amplitude $S$ \\
charged particle & the mass $m$ & the magnetic charge $q$ \\
symmetric spinning top & the moment of inertia $I_1$ & the angular momentum $M_3$\\
\hline
\end{tabular}

$~$\\

We have seen in this section that spin with finite inertia has an intrinsic precession mode (i.e. free precession or nutation).  The magnetic field causing this precession is the monopole magnetic field intrinsic to spin, i.e., the spin Berry curvature.  
   The frequency of the intrinsic precession mode, $\omega_0 \sim S/m_s$, is infinite when the inertia of spin $m_s$ is zero, but goes down to finite value when $m_s$ becomes non-zero.

As related works, we mention refs.\cite{Zhu04, Nussinov05, Zhu09, Fransson08, Fransson08no2, Fransson08no3} where nutational motion of spin in  Josephson junctions or tunnel junctions between ferromagnets was discussed by examining the Landau-Lifshitz-Gilbert equation with time-dependent Gilbert damping coefficient.

\section{Behavior of spin with spatial inhomogeneity and under time-dependent  magnetic field}\label{sec: phenomena}

We have so far assumed spatially homogeneous configuration of spin under time-independent magnetic field.   Let us include spatial variation of spin and time-dependence of applied magnetic field.  
The most typical behavior of spin under time-dependent magnetic field is the resonance phenomenon.  In ref.\cite{Olive12}, it is shown that spin with finite inertia $m_s$ has a resonance peak at the intrinsic frequency $\omega_0 \sim S/m_s$.  We discuss other behaviors of spin, such as spin wave, domain wall motion.  We also discuss an unusual behavior of spin under a large and time-dependent magnetic field.

\subsection{Spin waves}\label{sec:spin wave}

\begin{figure}[tbp]
\begin{center}
\includegraphics[width=150mm]{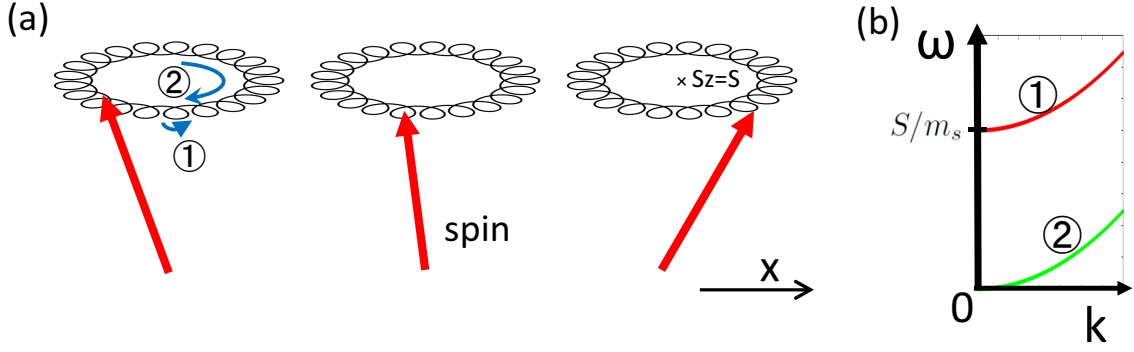}
\caption{\footnotesize{(a) The trajectory of a spin, which fluctuates around $S_z=S$ point (this fluctuation propagates in space as spin wave).  In the figure, two fluctuation modes are superimposed.   One is the usual gapless mode  rotating in clockwise seen from the positive $z$-direction (denoted by $\textcircled{2}$), while the other is a new gapful mode rotating counterclockwise (denoted by $\textcircled{1}$). (b) The dispersion relation for the two modes of spin wave.  Besides the usual gapless spin wave mode (in green color and denoted by $\textcircled{2}$), spin with inertia has a gapful spin wave mode (in red color and denoted by $\textcircled{1}$), which is nutation of spin propagating in space by the spin-spin exchange. See eq.\eqref{spin wave frequency}.}  }
\label{fig: spin wave}
\end{center}
\end{figure}

Let us consider spin wave solutions of spin with inertia, and see that there exists a new gapful spin wave mode with unusual handedness.  When we assume $\bm n = (n_x, n_y, 1)$ with $n_x, n_y \ll 1$, the equation of motion is\footnote{The assumption $n_x, n_y \ll 1$ means that we consider the problem in the tangent plane on $n_z=1$, so the constraint force in \eqref{real spin EOM} vanishes.}, from the action \eqref{spin effective action},
\begin{equation}
m_s \bm{\ddot n} = S\bm n \times \bm{\dot n} + JS^2\partial_i^2 \bm n
\label{spin wave equation}
\end{equation}
where we have set $\bm B=0$. Since the equation of motion is of second order in the time derivative, we have now two spin wave modes.   When we substitute a plane wave $n_x + i n_y = Ae^{-i(\omega t - k x)}$ to the equation of motion, with $A$ an arbitrary small constant,  we obtain the dispersion relation (Fig.\ref{fig: spin wave}),
\begin{equation}
\omega = \frac{1}{2}\frac{S}{m_s}\left(-1\pm \sqrt{1+4J m_s k^2}\right) \sim
\begin{cases}
 JSk^2 +  \mathcal O(k^4)\\
 -\left(\frac{S}{m_s}+JS k^2\right) + \mathcal O(k^4).
\end{cases}
\label{spin wave frequency}
\end{equation}
One is the usual spin wave mode and the other is the new spin wave mode, the latter of which is essentially the free precession of spin mediated in space by the interaction $J$ with neighboring spins.  We have expanded the dispersion relation \eqref{spin wave frequency} in powers of $k$ and cut higher order terms because we are in the framework of the derivative expansion in deriving  eq.\eqref{spin effective action}.  

The usual spin wave mode is clockwise seen from the positive $z$-direction, while the new spin wave mode is counterclockwise.  In fact, to the leading order of the dispersion relation \eqref{spin wave frequency}, the former obeys the equation of motion $S\bm {\dot n}=-J\partial_i^2\bm n\times \bm e_z=+Jk^2\bm n\times \bm e_z$, while the latter $m_s\bm {\ddot n}=-S\bm {\dot n}\times \bm e_z$, where $\bm e_z$ is the unit vector in $z$-direction.  Therefore, their handednesses are opposite.  The total angular momentum $\bm j^0$ in \eqref{angular momentum conservation} also rotates in counterclockwise for the new spin wave mode.

When the spin and the angular momentum align in the same direction, $\bm j^0=S\bm n$, the fluctuation of spin around the stable configuration is always clockwise.  This can be understood as follows (see e.g. ref.\cite{Onoda08}).  Let the stable configuration be $n_z=1$.  The Poisson commutation relation of the angular momenta, $[j^0_x,j^0_y]=j^0_z\cong S$, leads to that of spin, $[n_x, n_y]=1/S$.  Therefore, the $x$ and $y$ components of spin are canonically conjugate to each other.  Moreover, the Hamiltonian of spin expanded around the stable point is $H\cong D[(n_x)^2+(n_y)^2]$ with $D>0$.  These two facts, i.e. the canonical commutation relation of $n_x$ and $n_y$, and the form of the Hamiltonian, indicate that the fluctuation of spin around $n_z=1$ has the same dynamical structure as that of a one-dimensional harmonic oscillator with coordinate $x$ and its canonical momentum $p$ governed by the commutation relation $[x,p]=1$ and the Hamiltonian $H=D(x^2+p^2)$.
The trajectory of the harmonic oscillator is always clockwise in the phase space.  Therefore, the fluctuation of spin is also always clockwise.  However, when the inertia of spin is finite, the spin and the angular momentum generally do not align due to the additional contribution $m_s\bm n\times \bm{\dot n}$ in $\bm j^0$ \eqref{angular momentum conservation}, and moreover the Hamiltonian of spin such as eq.\eqref{H of spin} does not consist only of the potential term but also of the kinetic term.  Therefore, the fluctuation of spin with inertia is not restricted to clockwise motion.

\subsection{Domain wall}

\begin{figure}[tbp]
\begin{center}
\includegraphics[width=50mm]{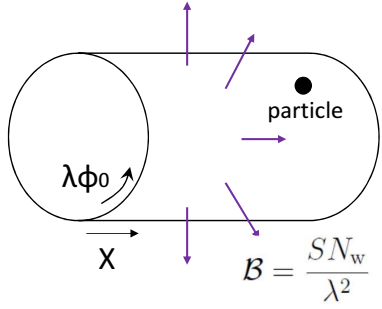}
\caption{\footnotesize{The collective coordinates $(X,\lambda\phi_0)$ of a domain wall can be regarded as coordinates of a particle on a cylinder.  This particle is subject to a magnetic field flux $\mathcal B=SN_{\rm w}/\lambda^2$ penetrating the cylinder perpendicularly.  The inertia of spin introduces the mass $M_{\rm w}=m_s N_{\rm w}/\lambda^2$ into this particle.}  }
\label{fig: particle on a cylinder}
\end{center}
\end{figure}

Let us see that a domain wall also has the free precession mode, corresponding to that of each spin.  
When easy-axis anisotropy $(KS^2/2)\cos^2\theta$ is included to the action \eqref{spin effective action}, the system has a static domain wall solution, $\cos\theta = \tanh[(z-X)/\lambda], \phi=\phi_0$ (we take $z$-axis as the wall normal) where $X$ and $\phi_0$ are integration constants, and $\lambda=\sqrt{J/K}$.  Promoting $X$ and $\phi_0$ to dynamical variables, substituting the domain wall solution into the action \eqref{spin effective action} and integrating it over space, we obtain the Lagrangian of $X$ and $\phi_0$:
\begin{equation}
L[X,\phi_0] = \frac{M_{\rm w}}{2}(\dot X^2 + \lambda^2\dot \phi_0^2) + \frac{SN_{\rm w}}{2\lambda}(\dot X\phi_0-X\dot\phi_0)
\label{L of domain wall}
\end{equation}
with $M_{\rm w}=m_s N_{\rm w}/\lambda^2$ and $N_{\rm w}=2\lambda A/a^3$ where $A$ is the cross sectional area of the domain wall (we discarded irrelevant constant terms in the Lagrangian).  We are going to show that this Lagrangian is of the same form as that of a charged particle on a cylinder with a constant magnetic field $\mathcal B=SN_{\rm w}/\lambda^2$ perpendicularly penetrating the surface of the cylinder.  Take an orthogonal coordinate frame $\bm x=(x,y)$ on a cylinder, with $y\sim y+2\pi\lambda$ the periodic direction.  (The directions $x,y$ and the direction outward normal to the surface of the cylinder make an orthogonal right-handed frame.)   Then, the gauge potential $\bm{\mathcal A}$ for the constant magnetic field $\mathcal B$ in the outward normal direction of the surface can be taken as $\mathcal A_x=\mathcal By/2, \mathcal A_y=-\mathcal Bx/2$, and the Lagrangian of the particle is 
\begin{equation}
L=\frac{m}{2}\bm{\dot x}^2 - \bm{\mathcal A}\cdot \bm{\dot x} 
= \frac{m}{2}(\dot x^2 + \dot y^2) + \frac{\mathcal B}{2}(\dot x y - x\dot y)
\end{equation}
Therefore, when we identify $(X,\lambda \phi_0)$ with $(x,y)$, the Lagrangian \eqref{L of domain wall} is that of a charged particle under a perpendicular magnetic field of flux $\mathcal B=SN_{\rm w}/\lambda^2$ (Fig.\ref{fig: particle on a cylinder}).  The magnetic field $\mathcal B$ for the particle originates in the spin Berry phase, and the mass $M_{\rm w}$ of the particle comes from
the inertia of spin in the action \eqref{spin effective action}.  

 As a result, the particle undergoes cyclotron motion (in other words, the domain wall undergoes the intrinsic free precession mode), where $X$ and $\lambda \phi_0$ oscillate, $X=r\cos(\omega_0 t), \lambda \phi_0=r\sin(\omega_0 t)$ with $r$ a constant, at frequency $\omega_0 =\mathcal B/M_{\rm w}= SN_{\rm w}/(\lambda^2 M_{\rm w})=S/m_s$.  This frequency corresponds to that of spin free precession discussed in section \ref{sec: spin and spinning top}.  
  When an external field is applied on this domain wall, the particle on a cylinder $(X,\lambda \phi_0)$ feels the corresponding force on it, and it moves  in the direction perpendicular to the force, accompanied by the cyclotron motion (nutation).  That is, the particle undergoes $\bm {\mathcal E}\times \bm {\mathcal B}$ drift motion, where $\bm {\mathcal E}$ is the force and $\bm {\mathcal B}$ is the magnetic field penetrating the cylinder $(X,\lambda \phi_0)$ perpendicularly.

The discussion above is well valid when the inertia $m_s$ is so large that the intrinsic frequency $\omega_0\sim S/m_s$ is less than the frequency of spin waves: $S/m_s <KS^2/\hbar 
$.   When $S/m_s$ exceeds this frequency scale, the dynamics of the domain wall can be no longer described only by the collective coordinates $X$ and $\phi_0$, and spin wave excitations should also be included into the dynamics.  

\subsection{Parallel shift of spin under time-dependent magnetic field}

In the spin nutation such as in Fig.\ref{fig: spin top particle}, spin moves  up and down parallelly to the given magnetic field, but this motion is oscillatory and the guiding center (i.e. the center of the nutation) moves only perpendicularly to the magnetic field, just as the usual Larmor precession.  
Is it possible that the guiding center motion becomes different from the usual Larmor precession and has a finite component of velocity parallel to the magnetic field, considering that the spin with inertia does not obey the usual Landau-Lifshitz equation?   
In this subsection, we see that when the magnetic field is time-dependent, the guiding center motion has finite component of velocity parallel to the time-derivative of the magnetic field. 
 
Bearing the particle analogy discussed in section \ref{sec: spin and spinning top} in mind, this parallel motion is most directly seen by looking at the behavior of a charged particle, located on a plane $(x,y)$ under a constant perpendicular magnetic field $\mathcal B$ and a time-dependent in-plane electric field $\bm {\mathcal E}(t)$ (recall that an electric field $\bm{\mathcal E}$ on the particle corresponds to a magnetic field $\bm B$ on spin).  The general solutions of the equation of motion is given by $(x, y)=(\xi,\eta)+(X, Y)$, such that  
\begin{equation}
m
\begin{pmatrix}
\ddot \xi \\
\ddot \eta
\end{pmatrix}
= \mathcal B
\begin{pmatrix}
0 & -1 \\
1 & 0 
\end{pmatrix}
\begin{pmatrix}
\dot \xi \\
\dot \eta
\end{pmatrix}.
\end{equation}
and
\begin{align}
\begin{pmatrix}
\dot X \\
\dot Y
\end{pmatrix}
= \frac{1}{\mathcal B}\frac{1}{1+\omega_c^{-2}(d^2/dt^2)}\begin{pmatrix}
\omega_c^{-1}d/dt & -1 \\
1 & \omega_c^{-1}d/dt
\end{pmatrix}
\begin{pmatrix}
\mathcal E_x \\
\mathcal E_y
\end{pmatrix},
~~~~\omega_c = \frac{\mathcal B}{m}.
\label{guiding center motion}
\end{align}
That is, we can divide the motion into the cyclotron motion $(\xi, \eta)$ and the guiding center motion $(X,Y)$ even when the electric field is time-dependent\footnote{As long as $\omega_c^{-1}d/dt<1$ and the operator $(1+\omega_c^{-2}d^2/dt^2)^{-1}$ is well-defined.}.  We can see that the guiding center motion $(X,Y)$ has a finite component of velocity parallel to the electric field, when the mass is finite and the electric field is time-dependent.  
As we saw in section \ref{sec: spin and spinning top}, the classical dynamics of spin with inertia is a spherical version of this planar problem, with the role of an electric field for the particle played by a magnetic field for spin, and therefore the essential features will be the same (especially when the nutation radius is small and the curvature of the sphere can be neglected): (i) the spin nutation and the guiding center motion are decoupled; (ii) the effect of the inertia on the guiding center motion is significant when the time rate of change of the applied field is comparable to the intrinsic frequency $S/m_s$.  

Therefore, the effect of the inertia is not only the superimposition of the nutational motion to the Larmor precession.  The inertia of spin gives an independent effect on the trajectory of the Larmor precession itself, cooperatively with the time-dependence of the applied field.    Even when the nutation radius is small and there are no signs of nutation in experimental data about the time-profile of magnetization, it never indicates that the inertia has no effects on the dynamics, especially when the applied magnetic field has very rapid time-dependence, such as ultrafast magnetization.  

\begin{figure}[tbp]
\includegraphics[width=165mm]{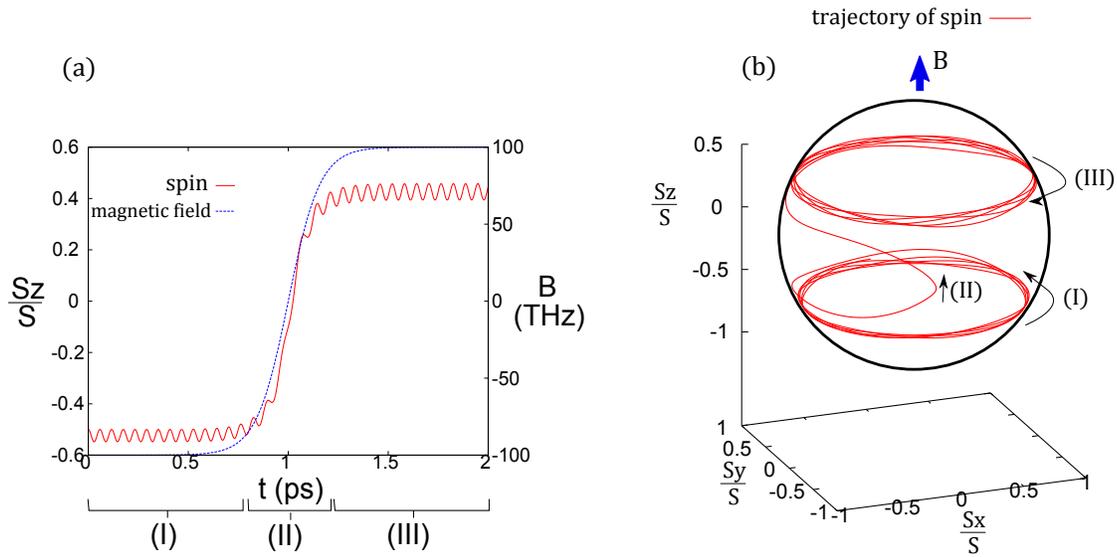}
\caption{\footnotesize{Response of spin with finite inertia to a time-dependent magnetic field $B(t)=B_0\tanh(f(t-t_0))$ in $z$-direction with $B_0=100{\rm THz}$ and $f=2\pi{\rm THz}=6.3 {\rm THz}$.   We take the spin amplitude as $S=\hbar$ and the inertia as $m_s/\hbar=0.1{\rm ps}/(2\pi)=0.016{\rm ps}$.  (a) Time profile of $S_z/S(=n_z)$ of spin and applied magnetic field $B$. (b) The trajectory of spin.  The time is divided into three domains I-III. In I, the magnetic field points in the negative $z$-direction and spin precesses counterclockwise seen from the positive $z$-direction.  In II, the magnetic field switches to positive $z$-direction and the $z$-component of spin, $S_z/S$, also changes accordingly.  In III, the magnetic field points in positive $z$-direction and spin precesses clockwise. }}
\label{fig: reaction to time-dependent magnetic field}
\end{figure}

We illustrate this fact by numerical calculation (Fig.\ref{fig: reaction to time-dependent magnetic field}).  We fix a magnetic field in the $z$-direction.  First, we make the magnetic field in the negative $z$-direction, then change it into the positive $z$-direction.  We can see that the spin moves parallelly (i.e. in the positive $z$-direction) to the time derivative of the magnetic field when the magnetic field is changing\footnote{The switching of the sign of the magnetic field from negative to positive is not essential for the shift of the spin in $z$-direction.  Only the time-dependence of the magnetic field is essential.}.  This phenomenon is similar to usual spin flip by the Gilbert damping, but is different in that we do not have to vary the direction of  the magnetic field and moreover the time change rate of the magnetic field can be arbitrary (as long as our derivative expansion is valid) and is irrelevant to the damping coefficient.  

\begin{figure}
\begin{minipage}[htbp]{.47\textwidth}
\vspace{3cm} 
\includegraphics[width=70mm]{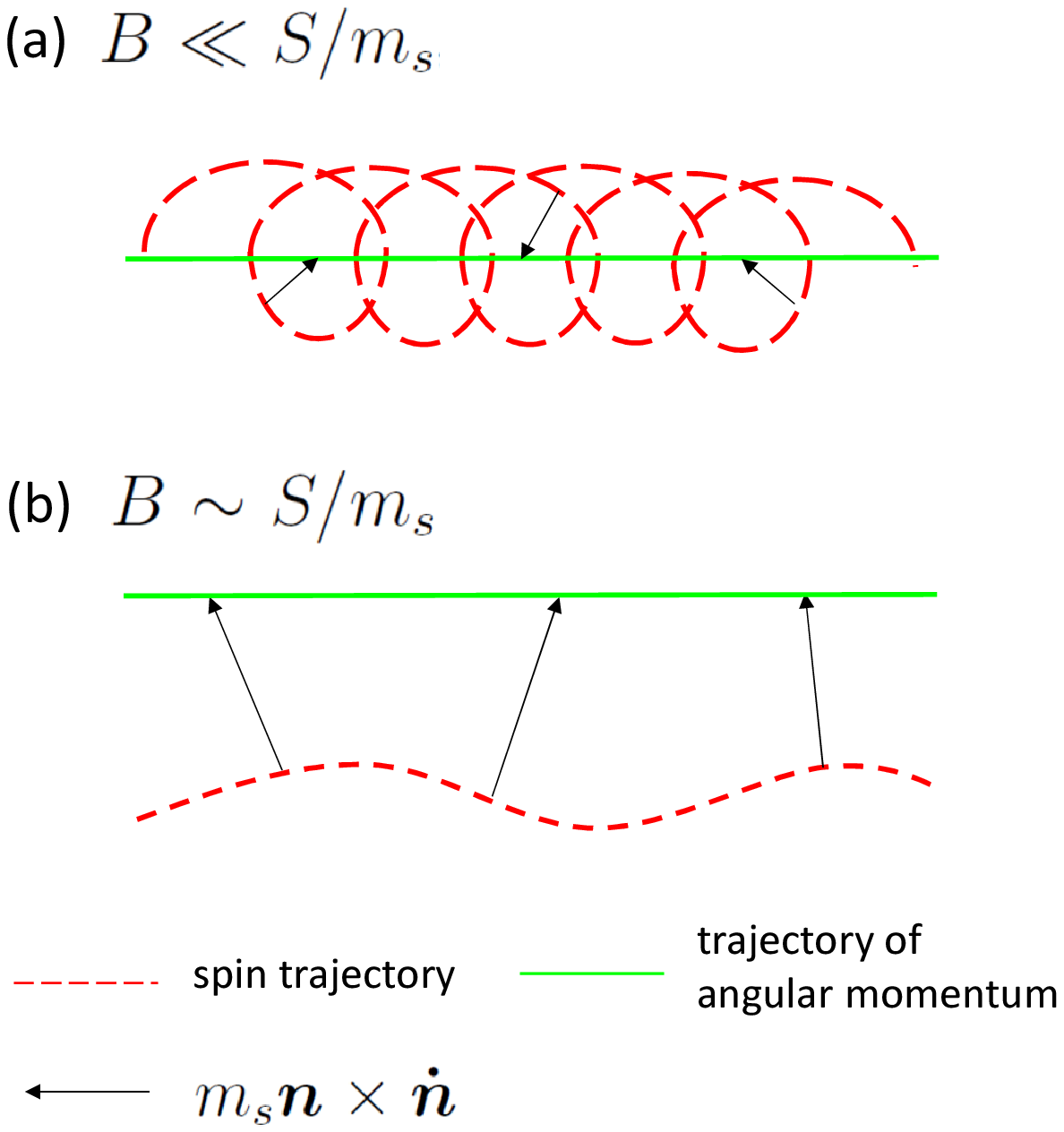}
\caption{\footnotesize{Trajectories of spin $S\bm n$ and angular momentum $\bm j^0=S\bm n + m_s\bm n \times \bm {\dot n}$. (a) When $B\ll S/m_s$, spin nutation forms circle shape.  The additional component $m_s\bm n \times \bm {\dot n}$ points inward to the center of the nutation, and the trajectory of the angular momentum sits within that of spin.  (b) When $B\sim S/m_s$, spin nutation forms ripple shape.  The additional component $m_s\bm n \times \bm {\dot n}$ points to one side of spin trajectory.  As the result, the trajectories of spin and the angular momentum deviate.  }}
\label{fig: contribution of orbital angular momentum}
\end{minipage} 
\hfill
\begin{minipage}[htbp]{.47\textwidth}
\vspace{3cm} 
\includegraphics[width=80mm]{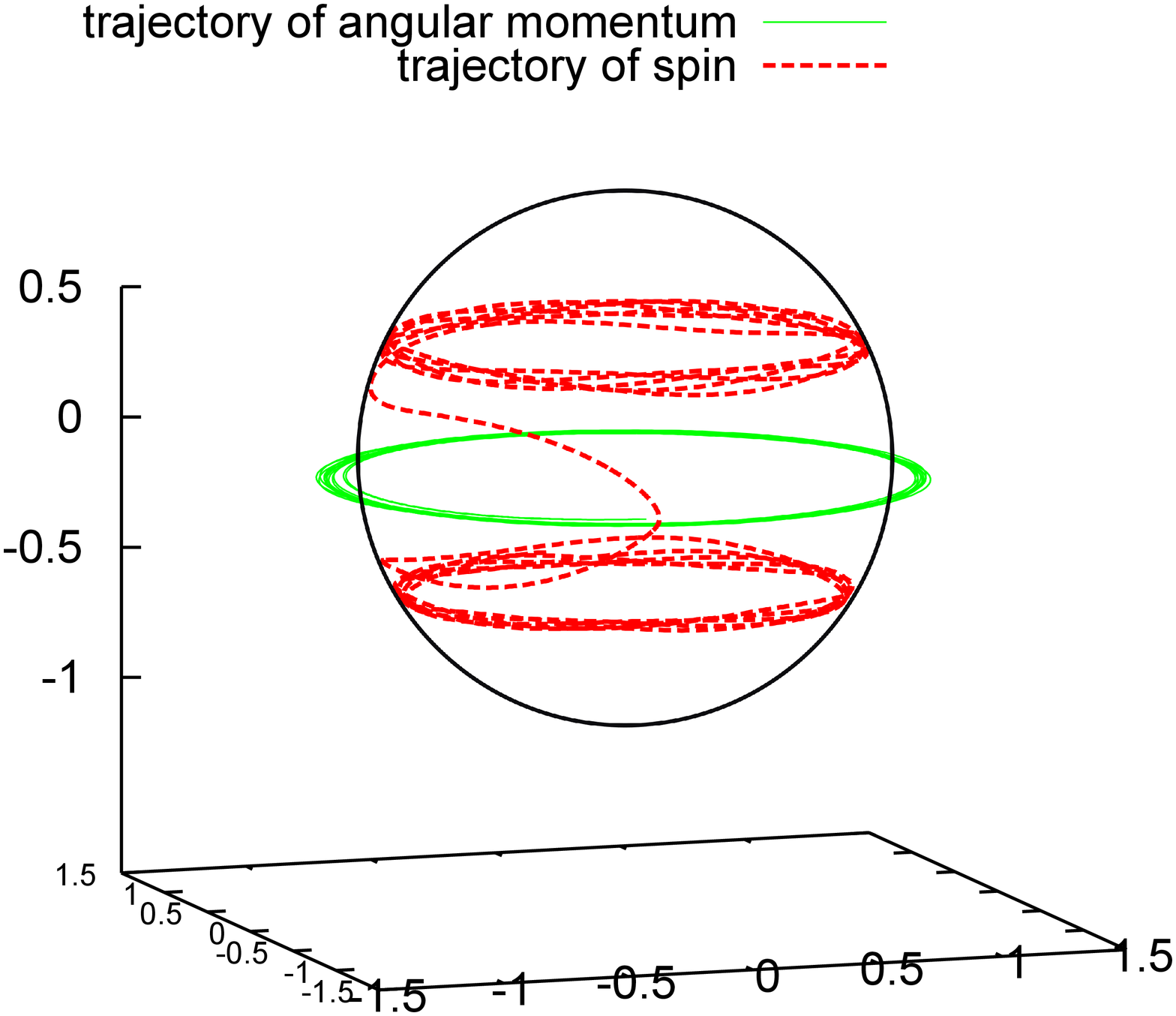}
\caption{\footnotesize{Replot of the trajectory of spin in Fig.\ref{fig: reaction to time-dependent magnetic field} (from slightly a different view angle), together with that of the angular momentum.  The magnetic field is $B\sim 100{\rm THz}$ and $S/m_s\sim 20\pi{\rm THz}=62.8{\rm THz}$, corresponding to the case (b) in Fig.\ref{fig: contribution of orbital angular momentum}.  Although the $z$-component of spin changes in time, that of the angular momentum is constant in motion. The axes are in unit of $S$. }}
\label{fig: trajectories of spin and angular momentum}
\end{minipage}
\end{figure}

This unusual behavior of spin can be simply understood when we look at the angular momentum rather than spin itself.  As in \eqref{angular momentum conservation}, the angular momentum of spin with finite inertia is no longer $S\bm n$ but has additional non-adiabatic contribution $m_s \bm n \times \bm{\dot n}$.  When the Larmor frequency $B$ is much smaller than the intrinsic frequency (nutation frequency), $B\ll S/m_s$, then the vector $m_s \bm n \times \bm{\dot n}$ points inward to the center of the nutation circle, and the trajectory of the angular momentum of spin is overlapped with that of spin (Fig.\ref{fig: contribution of orbital angular momentum}(a)).  On the other hand, when $B\sim S/m_s$, nutation does not form circles but rather ripples on the trajectory of spin.  In this case, the vector $m_s \bm n \times \bm{\dot n}$ points in almost the same direction at each time.  As the result, the trajectory of the angular momentum deviates from that of spin (Fig.\ref{fig: contribution of orbital angular momentum}(b)).  Fig.\ref{fig: trajectories of spin and angular momentum} is a replot of the trajectory of spin in Fig.\ref{fig: reaction to time-dependent magnetic field} together with that of the angular momentum.  The vector $m_s \bm n \times \bm{\dot n}$ points in the positive $z$-direction in the time domain I (described in Fig.\ref{fig: reaction to time-dependent magnetic field}), and in the negative $z$-direction in the time domain III, resulting in the conservation of the $z$-component of the angular momentum.  

This parallel shift of spin in the direction of the time derivative of the magnetic field can be understood also from a viewpoint of a spinning top, where the magnetic field on spin corresponds to the gravitational field on the top.  The time-variation of the gravitational field can be realized fictitiously by putting the top in an elevator accelerated in the negative $z$-direction.  There, the top feels, besides the real gravitational field in the negative $z$-direction, a fictitious  inertial force in the positive $z$-direction due to the acceleration.  This inertial force in the positive $z$-direction makes the top stand up in the positive $z$-direction, which explains the shift of spin in $z$-direction.  Thus, the inertia of spin yields an inertial force, literally.  In this respect, the connection between the inertia of spin and the inertial effect on spin \cite{Hehl90, Matsuo11} will also be interesting.

\section{Summary and discussion}

We examined the inertia of spin in metallic ferromagnets, which arises in the derivative expansion of spin effective action and represents non-adiabatic contribution from environmental degrees of freedom.  We derived an explicit expression of spin inertia in an $sd$ model.  The equivalence between the dynamics of spin with inertia, spinning top and a charged particle on a sphere was explained.  The behavior of spin with spatially inhomogeneous configuration or under time-dependent magnetic field is studied.  The finite inertia has mainly two effects: the superimposition of the nutation on the usual Larmor precession and the parallel shift of the Larmor precession.  


As explained several times in this paper, most of unusual behavior of spin with finite inertia originates in the non-adiabatic component $m_s\bm n \times \bm {\dot n}$ in the relation between spin and the angular momentum, $\bm j^0=S\bm n+ m_s\bm n \times \bm {\dot n}$.  Due to this difference between them, spin acquires a kind of separate freedom from the angular momentum, whose behavior is relatively restricted due to the conservation law.  As the result, spin undergoes free precession or nutation discussed in section \ref{sec: spin and spinning top}, and furthermore parallel shift along the direction of the time derivative of applied magnetic fields, discussed in section \ref{sec: phenomena}.  Since one of the main purposes of spintronics is understanding and application of diverse cooperative phenomena between localized spins and their environmental degrees of freedom, researches beyond the adiabatic limit will be well-motivated for this spirit of spintronics;  in the adiabatic limit, the environmental degrees of freedom are rather obedient to the localized spins.  In this paper, we estimated that in metallic ferromagnets such non-adiabaticity is effective for   sub-picosecond dynamics, i.e. ultrafast magnetization.  For broader applications, physical situations where non-adiabaticity is effective for sub-nanosecond dynamics are also desirable.  

Apart from the context of effective dynamics (i.e. the dynamics including the effects of environmental degrees of freedom), the inertial term, together with the spin Berry phase term, can be present in the dynamics of the order parameters in systems whose ground states can be intermediate states between ferromagnet and antiferromagnet (see e.g.\cite{Kawaguchi}).  The discussion of this paper will be also applicable to such systems.

\section{Acknowledgment}
T. K. would like to thank Aron Beekman, Yan-Ting Chen, Hideo Kawaguchi, Se Kwon Kim, Nguyen Thanh Phuc, Henri Saarikoski, members in Eiji Saitoh group in Tohoku university, and other people for helpful comments.   

\begin{appendix}

\section{spin effective action}\label{sec: calculation of spin effective action}

In this appendix, we calculate the spin effective action by integrating out a conduction electron coupled to spin through the $sd$ interaction, and derive the expression of spin inertia \eqref{expression of inertia}.  We always make summation over repeated indices unless otherwise stated.  

The action of a conduction electron, represented by field operators $c$ and $\bar c$, coupled to localized spin $\bm n$ is 
\begin{equation}
\mathcal S_{\rm e}[c, \bar c,\bm n]=\int d^3x dt~\bar c \left(i\hbar \partial_t + \frac{\hbar^2\partial_i^2}{2m}
+\epsilon_F + g_{\rm sd}\bm n \cdot \bm \sigma \right)c.
\label{action of conducting fermion}
\end{equation}
with $g_{\rm sd}$ the $sd$ coupling constant.  What we calculate in this appendix is the contribution to spin effective action from the conduction electron, $\Delta \mathcal S_{\rm eff}[\bm n]$, defined as 
\begin{equation}
\exp\left(\frac{i}{\hbar}\Delta \mathcal S_{\rm eff}[\bm n]\right) \equiv \int \mathcal D\bar c \mathcal D c~ \exp\left(\frac{i}{\hbar}\mathcal S_{\rm e}[c,\bar c, \bm n]\right).
\label{def of Delta Seff}
\end{equation}
Adding this $\Delta \mathcal S_{\rm eff}[\bm n]$ to the original spin action 
\begin{equation}
\mathcal S_{\rm s}[\bm n] = \int \frac{d^3x}{a^3}\left[S_l \dot \phi(\cos\theta-1) - J_lS_l^2(\partial_i \bm n)^2\right],
\end{equation}
where $\theta$ and $\phi$ are the polar angles of the unit vector $\bm n$, we obtain the total spin effective action $\mathcal S_{\rm eff}[\bm n] = \mathcal S_{\rm s}[\bm n]+\Delta \mathcal S_{\rm eff}[\bm n]$.  

First, we diagonalize the $sd$ coupling in \eqref{action of conducting fermion}.  In order for that purpose, it is convenient and essential to employ the unitary-transformed electron field $a$ such that $c(\bm x,t)=U(\bm x,t)a(\bm x,t)$.  Here, the $SU(2)$ matrix $U$ is defined so that 
\begin{equation}
U^{\dagger} (\bm n \cdot \bm \sigma)U = \sigma_3.
\label{definition of U}     
\end{equation}
As such, $U$ has indefiniteness about the right $U(1)$ gauge transformation, $U\rightarrow U\exp(i\chi(\bm x,t)\sigma_3)$ with $\chi(\bm x,t)$ an arbitrary  function.  [Due to the defining relation $c=Ua$ of $a$, this unitary transformation corresponds to that of $a$, i.e. $a\rightarrow \exp(-i\chi(\bm x,t)\sigma_3)a$.  The original electron field $c$ remains invariant.]  One choice of $U$ is $U=\bm m\cdot \bm \sigma$ with $\bm m=(\sin\frac{\theta}{2}\cos\phi,\sin\frac{\theta}{2}\sin\phi,\cos\frac{\theta}{2})$.  
Geometrically, $U$ rotates the electron spin in the positive $z$-direction, $|\uparrow \rangle$, to the direction of $\bm n$.  For the choice $U=\bm m\cdot \bm \sigma$ above, $U|\uparrow\rangle = |\bm n\rangle$ with $|\bm n\rangle\equiv \cos(\theta/2)|\uparrow\rangle + e^{i\phi}\sin(\theta/2)|\downarrow\rangle$.  The indefiniteness of $U$, $U\rightarrow U\exp(i\chi\sigma_3)$, corresponds to the phase rotation of $|\bm n\rangle$, i.e. $|\bm n\rangle \rightarrow \exp(i\chi)|\bm n\rangle$.  
The $sd$ coupling becomes diagonal in the $a$-frame, $\bar c (\bm n \cdot \bm \sigma)c = \bar a \sigma_3 a$ but, in compensation for the $SU(2)$ gauge transformation by $U$, there appears a spin gauge field $A_\mu\equiv -iU^{\dagger}\partial_\mu U=(\bm m\times \partial_\mu \bm m)\cdot \bm \sigma$ in the derivative terms, $\partial_\mu c=U(\partial_\mu+iA_\mu)a$. 

Then, the action \eqref{action of conducting fermion} becomes \cite{Tatara94}
\begin{align}
\mathcal S_{\rm e}=&\int d^3x dt~\bar a \left(i\hbar (\partial_t+iA_t) + \frac{\hbar^2(\partial_i+iA_i)^2}{2m}
+\epsilon_F + g_{\rm sd} \sigma_3 \right)a
\notag \\
=& \mathcal S_0[a, \bar a] + \int d^3x dt \left[
A_t^a (-\hbar \bar a\sigma^a a) + A_i^a \left(\frac{i\hbar^2}{2m}\bar a \sigma_a \overleftrightarrow{\partial}_i a\right) + A_i^a A_i^a \left(-\frac{\hbar^2}{2m}\bar a a\right)
\right].
\label{action of conducting fermion2}
\end{align}
Here $\mathcal S_0[a, \bar a]$ is 
\begin{equation}
\mathcal S_0[a, \bar a]\equiv \int d^3x dt~\bar a \left(i\hbar \partial_t + \frac{\hbar^2\partial_i^2}{2m}
+\epsilon_F + g_{\rm sd} \sigma_3 \right)a,
\label{S0}
\end{equation}
and $A_\mu =A_\mu^a \sigma^a~ (a=1,2,3)$ is the spin gauge field (make distinction between the electron field $a$ and the $SU(2)$ index$~^a$), which is, in terms of the polar angles of $\bm n$,   
\begin{equation}
\bm A_\mu =\frac{1}{2}
\begin{pmatrix}
-\partial_\mu\theta\sin\phi - \sin\theta \cos\phi\partial_\mu\phi \\
\partial_\mu\theta\cos\phi - \sin\theta\sin\phi\partial_\mu\phi \\
(1-\cos\theta)\partial_\mu\phi
\end{pmatrix}
\label{spin gauge field}
\end{equation} 
where $(\bm A_\mu)^a=A_\mu^a$.  Because $A_\mu$ is proportional to the first order derivative of $\bm n$, the derivative expansion of $\Delta \mathcal S_{\rm eff}$ in powers of $\partial_\mu \bm n$ corresponds to that of $A_\mu$: 
\begin{align}
\frac{i}{\hbar}\Delta \mathcal S_{\rm eff}[\bm n] =& 
\frac{i}{\hbar}\int d^4x \left[
A_t^a (-\hbar \langle \bar a\sigma^a a\rangle) + A_i^a \frac{i\hbar^2}{2m}\langle \bar a \sigma_a \overleftrightarrow{\partial}_i a\rangle + A_i^a A_i^a \left(-\frac{\hbar^2}{2m}\langle \bar a a\rangle \right)
\right] \notag \\
&+ \frac{1}{2}\left(\frac{i}{\hbar}\right)^2 \int d^4x d^4x'
\Big[ 
A_t^a(x) A_t^b(x')\hbar^2 \langle \bar a \sigma^a a(x)\bar a \sigma^b a(x') \rangle
\notag \\
&+ A_i^a(x)A_j^b(x')\left(\frac{i\hbar^2}{2m}\right)^2 \langle \bar a\sigma^a \overleftrightarrow{\partial_i} a(x) \bar a\sigma^b \overleftrightarrow{\partial_j} a(x') \rangle 
\notag \\
&+ A_t^a(x)A_i^b(x')\left(-\frac{i\hbar^3}{m}\right)
\langle \bar a \sigma^a a(x)\bar a \sigma^b \overleftrightarrow{\partial_i}a(x')\rangle
\Big] + \mathcal O(A_\mu^3)
\label{expansion in powers of A}
\end{align}
where time-ordered quantum expectation values are taken about $\mathcal S_0$,
\begin{equation}
\langle \cdots \rangle = \int \mathcal D \bar a \mathcal D a ~ (\cdots) e^{\frac{i}{\hbar}\mathcal S_0[a, \bar a]},
\end{equation}
and we take only the connected diagrams to evaluate the quantum expectation values.    
Because the non-perturbed action $\mathcal S_0$ \eqref{S0} is isotropic in space, $\langle \bar a \sigma^a \overleftrightarrow{\partial}_i a\rangle$ and $\langle \bar a \sigma^a a(x)\bar a \sigma^b \overleftrightarrow{\partial_i}a(x')\rangle$ vanish.  
We define the notation for correlation functions as 
\begin{equation}
\rho^{ab}(x,x') \equiv i \langle \bar a \sigma^a a(x)\bar a \sigma^b a(x') \rangle,~~
\chi^{ab}_{ij}(x,x') \equiv i \langle \bar a\sigma^a \overleftrightarrow{\partial_i} a(x) \bar a\sigma^b \overleftrightarrow{\partial_j} a(x') \rangle.
\end{equation}
Using the Fourier components, we expand these non-local quantities.  For example, 
\begin{align}
\rho^{ab}(x,x') =& \int \frac{d^4q}{(2\pi)^4}~e^{i q\cdot (x- x')}\rho^{ab}(q)
\notag \\
=&  \delta^{(4)}(x-x')\rho^{ab}|_{q=0} 
+ \frac{1}{2i}\partial_\mu \delta^{(4)}(x-x') \frac{\partial \rho^{ab}}{\partial q^\mu}|_{q=0}
+\cdots.
\label{expansion of nonlocal quantity}
\end{align}
When we substitute this into the expansion \eqref{expansion in powers of A}, the terms other than the first term in \eqref{expansion of nonlocal quantity} increase the order of the derivative on $\bm n$ after partial integration, e.g. $A_t^a(x)A_t^b(x')\partial_\mu \delta^{(4)}(x-x') \rightarrow -\partial_\mu [A_t^a(x) A_t^b(x')] \delta^{(4)}(x-x')$.  Therefore, we take only the first term in \eqref{expansion of nonlocal quantity} as long as we are interested in at most $\mathcal O(\partial_\mu \bm n)^2$-terms.  
After all, $\Delta \mathcal S_{\rm eff}[\bm n]$ is, to $(\partial_\mu \bm n)^2$-order,
\begin{equation}
\Delta \mathcal S_{\rm eff}[\bm n] = 
\int d^4x \left[
(-\hbar \langle \bar a\sigma^a a\rangle)A_t^a  + \left(-\frac{\hbar^2}{2m}\langle \bar a a\rangle \delta_{ab}\delta_{ij} - \frac{\hbar^3}{8m^2}{\rm Re}\chi^{ab}_{ij}|_{q=0} \right)A_i^a A_j^b
+ \left(\frac{\hbar}{2}{\rm Re}\rho^{ab}|_{q=0}\right)A_t^a A_t^b
\right]
\label{Delta Seff}
\end{equation}
(the imaginary parts of $\chi^{ab}_{ij}$ and $\rho^{ab}$ contribute to damping and are neglected here).  We calculate each term in the following.  

Firstly, we calculate $\langle \bar a\sigma^a a\rangle$.  The time-ordered Green's function for $\mathcal S_0$ \eqref{S0} is 
\begin{align}
 -i\langle a_\alpha(x) \bar a_\beta(x')\rangle 
=& \delta_{\alpha\beta}\int \frac{d^3kd\omega}{(2\pi)^4}
e^{i\bm k\cdot(\bm x-\bm x')-i\omega (t-t')}g_{\bm k,\omega,\alpha} ~~(\mbox{no summation over}~\alpha), \notag \\
&g_{\bm k,\omega,\alpha} \equiv \frac{1}{\omega-\omega_{\bm k\alpha}+i\delta_\alpha},
\label{Green function of S0}
\end{align} 
(the indices $\alpha,\beta,\cdots = \pm$ correspond to spin up or down) with 
\begin{equation}
\hbar \omega_{\bm k\pm} \equiv \frac{\hbar^2\bm k^2}{2m}-\epsilon_F \mp g_{\rm sd}, ~~
\delta_\alpha = {\rm sgn}(\omega_{\bm k\alpha})\times 0, 
\end{equation}
where 0 denotes a positive infinitesimal.  We define a $2\times 2$ matrix $g_{\bm k,\omega}$ by $(g_{\bm k,\omega})_{\alpha\beta} \equiv \delta_{\alpha\beta}g_{\bm k,\omega,\alpha}$ (no summation over $\alpha$), or, 
\begin{equation}
g_{\bm k,\omega} \equiv \frac{1}{2}(g_{\bm k,\omega,+}+g_{\bm k,\omega,-} + \sigma^3(g_{\bm k,\omega,+}-g_{\bm k,\omega,-})).
\end{equation}
Then, 
\begin{align}
\langle \bar a\sigma^a a\rangle =& -i\int \frac{d^3kd\omega}{(2\pi)^4}e^{i\omega0}{\rm tr}[\sigma^a g_{\bm k,\omega}]
= -i\delta^a_3\sum_{\pm}(\pm)\int \frac{d^3kd\omega}{(2\pi)^4}
\frac{e^{i\omega0}}{\omega-\omega_{\bm k\pm}+i\delta_\pm}
\notag \\
=& \delta^a_3\sum_{\pm}(\pm)\int \frac{d^3k}{(2\pi)^3}\theta(-\omega_{\bm k\pm})
= \frac{k_{F+}^3-k_{F-}^3}{6\pi^2}\delta^a_3,~~~\frac{\hbar^2k_{F\pm}^2}{2m}\equiv\epsilon_F\pm g_{\rm sd},
\end{align}
where $\theta(x)=1$ for $x\geq0$ and zero otherwise.

Secondly, the number density of the conducting electrons $n\equiv \langle \bar a a\rangle$ is given by $n= (k_{F+}^3+k_{F-}^3)/(6\pi^2)$.   


Thirdly, we calculate ${\rm Re}\chi^{ab}_{ij}|_{q=0}$.  Using 
\begin{equation}
{\rm tr}[\sigma^a g_{\bm k,\omega}\sigma^b g_{\bm k + \bm q, \omega+\Omega}] 
= \sum_\pm \left[(\delta_{ab}-\delta_{a3}\delta_{b3} \mp i\varepsilon_{ab3})g_{\bm k,\omega,\pm}g_{\bm k+\bm q,\omega+\Omega,\mp} 
+ \delta_{a3}\delta_{b3}g_{\bm k,\omega,\pm}g_{\bm k+\bm q,\omega+\Omega,\pm}\right]
\end{equation}
and 
\begin{equation}
-i\int \frac{d\omega}{2\pi}g_{\bm k,\omega,\alpha}g_{\bm k+\bm q,\omega+\Omega,\beta} = \frac{\theta(\omega_{\bm k\alpha})\theta(-\omega_{\bm k+\bm q\beta})}{\omega_{\bm k+\bm q\beta}-\omega_{\bm k\alpha}-\Omega+i0}
- \frac{\theta(-\omega_{\bm k\alpha})\theta(\omega_{\bm k+\bm q\beta})}{\omega_{\bm k+\bm q\beta}-\omega_{\bm k\alpha}-\Omega-i0},
\end{equation}
we obtain 
\begin{align}
\chi^{ab}_{ij}(\bm q,\Omega) =& -i\int  \frac{d^3kd\omega}{(2\pi)^4}(2k_i+q_i)(2k_j+q_j){\rm tr}[\sigma^a g_{\bm k,\omega}\sigma^b g_{\bm k+\bm q,\omega+\Omega}]
\notag \\
=&\sum_\pm\int \frac{d^3k}{(2\pi)^3}(2k_i+q_i)(2k_j+q_j)
\notag \\
&\times\Big\{(
\delta_{ab}-\delta_{a3}\delta_{b3}\mp i\varepsilon_{ab3})
\left[\frac{\theta(\omega_{\bm k\pm})\theta(-\omega_{\bm k+\bm q\mp})}{\omega_{\bm k+\bm q\mp}-\omega_{\bm k\pm}-\Omega+i0}
- \frac{\theta(-\omega_{\bm k\pm})\theta(\omega_{\bm k+\bm q\mp})}{\omega_{\bm k+\bm q\mp}-\omega_{\bm k\pm}-\Omega-i0}\right]
\notag \\
& ~~~~~~~+ \delta_{a3}\delta_{b3}
\left[\frac{\theta(\omega_{\bm k\pm})\theta(-\omega_{\bm k+\bm q\pm})}{\omega_{\bm k+\bm q\pm}-\omega_{\bm k\pm}-\Omega+i0}
- \frac{\theta(-\omega_{\bm k\pm})\theta(\omega_{\bm k+\bm q\pm})}{\omega_{\bm k+\bm q\pm}-\omega_{\bm k\pm}-\Omega-i0}\right]
\Big\}.
\label{chi intermediate}
\end{align}
We are interested here only in the real part, ${\rm Re}\chi^{ab}_{ij}$.  Using 
\begin{equation}
\frac{1}{x\pm i0}= {\rm P}\frac{1}{x}\mp i \pi \delta(x)~~~{\rm and}~~~
\theta(x)\theta(-y)-\theta(-x)\theta(y) = \theta(x) - \theta(y) = \theta(-y)-\theta(-x), 
\end{equation}
then,
\begin{align}
{\rm Re}\chi^{ab}_{ij}(\bm q,\Omega) 
=&\sum_\pm\int \frac{d^3k}{(2\pi)^3}(2k_i+q_i)(2k_j+q_j)
\notag \\
&\times\Big\{(
\delta_{ab}-\delta_{a3}\delta_{b3}){\rm P}
\frac{\theta(-\omega_{\bm k+\bm q\mp})-\theta(-\omega_{\bm k\pm})}{\omega_{\bm k+\bm q\mp}-\omega_{\bm k\pm}-\Omega}
+ \delta_{a3}\delta_{b3}
{\rm P}\frac{\theta(-\omega_{\bm k+\bm q\pm})-\theta(-\omega_{\bm k\pm})}{\omega_{\bm k+\bm q\pm}-\omega_{\bm k\pm}-\Omega}
\notag \\
& ~~~~~~\pm \varepsilon_{ab3}\pi[\theta(-\omega_{\bm k+\bm q\mp})-\theta(-\omega_{\bm k\pm})] \delta(\omega_{\bm k+\bm q\mp}-\omega_{\bm k\pm}-\Omega)
\Big\}.
\label{Rechi intermediate}
\end{align}
Let us calculate the value of each component of ${\rm Re}\chi^{ab}_{ij}|_{q=0}$.  First, 
\begin{align}
{\rm Re}\chi^{11}_{ij}|_{q=0}={\rm Re}\chi^{22}_{ij}|_{q=0}=&
\sum_\pm\int \frac{d^3k}{(2\pi)^3}4k_i k_j(\pm)\frac{\hbar}{2g_{\rm sd}}[\theta(-\omega_{\bm k\mp})-\theta(-\omega_{\bm k\pm})]
\notag \\
=& -\frac{\hbar}{g_{\rm sd}}\frac{2(k_{F+}^5-k_{F-}^5)}{15\pi^2}\delta_{ij}.
\end{align}
These components do not depend on the order of taking the $\bm q\rightarrow 0$ and $\Omega \rightarrow 0$ limit.  On the other hand, the value of the component ${\rm Re}\chi^{33}_{ij}$ depends on the order:  From eq.\eqref{Rechi intermediate}, it is clear that when we take the $\bm q \rightarrow 0$ limit first, ${\rm Re}\chi^{33}_{ij}$ vanishes,  while when we take the $\Omega\rightarrow 0$ limit first and then take the $\bm q\rightarrow 0$ limit, it gives $0/0$ and, using the L'Hopital's rule,  
\begin{equation}
\lim_{\bm q\rightarrow 0}\lim_{\Omega\rightarrow 0}{\rm Re}\chi^{33}_{ij}= 
\sum_\pm\int \frac{d^3k}{(2\pi)^3}4k_i k_j(-1)\delta(-\omega_{\bm k\pm})
= -\frac{4mn}{\hbar}\delta_{ij}.
\label{chi33}
\end{equation}
The order of the $\bm q\rightarrow 0, \Omega\rightarrow 0$ limits can be simply determined by the isotropy of the internal space (i.e. the space where the vector $\bm n$ is defined).  For example,  the combination $(A_i^1)^2 + (A_i^2)^2 = (1/4)(\partial_i \bm n)^2$ (see eq.\eqref{spin gauge field}) is isotropic (i.e. does not depend on the rotation $\bm n\rightarrow R\bm n$ with a constant $SO(3)$ matrix $R$), but $(A_i^3)^2= (1/4)(1-\cos\theta)^2(\partial_i\phi)^2$ is not an isotropic quantity.  When we employ $\lim_{\Omega \rightarrow 0}\lim_{\bm q\rightarrow 0}{\rm Re}\chi^{33}_{ij}=0$ as the value of ${\rm Re}\chi^{33}_{ij}|_{q=0}$, then the $(A_i^3)^2$-term appears in $\Delta \mathcal S_{\rm eff}$ \eqref{Delta Seff} due to the nonzero $\langle \bar a a\rangle$-term.  On the other hand, when we employ eq.\eqref{chi33} as the value of ${\rm Re}\chi^{33}_{ij}|_{q=0}$, then the $(A_i^3)^2$-term vanishes in $\Delta \mathcal S_{\rm eff}$ due to the cancellation with the $\langle \bar a a\rangle$-term.  Therefore, we should employ eq.\eqref{chi33} as the value of ${\rm Re}\chi^{33}_{ij}|_{q=0}$.\footnote{
The order of $\bm q\rightarrow 0, \Omega\rightarrow 0$ limits can be determined also by the $U(1)$ gauge symmetry $U\rightarrow U\exp(i\chi\sigma_3),~a\rightarrow \exp(-i\chi\sigma_3)a$ described below eq.\eqref{definition of U}.  Under this gauge transformation, the spin gauge field $A_\mu=A_\mu^a\sigma^a$changes as 
\begin{equation}
\begin{pmatrix}
A_\mu^1 \\
A_\mu^2
\end{pmatrix}
\rightarrow
\begin{pmatrix}
\cos2\chi & -\sin2\chi \\
\sin2\chi & \cos2\chi
\end{pmatrix}
\begin{pmatrix}
A_\mu^1 \\
A_\mu^2
\end{pmatrix}
,~~A_\mu^3\rightarrow A_\mu^3+\partial_\mu\chi.
\end{equation}   
The effective action $\Delta \mathcal S_{\rm eff}$ \eqref{Delta Seff} should be invariant under this gauge transformation.  Therefore, $(A_i^1)^2+(A_i^2)^2$-term is allowed but $(A_i^3)^2$-term is not.  
}
  

For non-diagonal elements, it is clear that ${\rm Re}\chi^{12}_{ij}|_{q=0}=0$ regardless of the order of the $\bm q\rightarrow 0$ and $\Omega\rightarrow 0$ limits.


Finally, let us calculate ${\rm Re} \rho^{ab}|_{q=0}$.  The calculation is almost the same as that of ${\rm Re} \chi^{ab}_{ij}|_{q=0}$:  just drop the factor $-(2k_i+q_i)(2k_j+q_j)$ from eq.\eqref{chi intermediate}.  Then, 
\begin{align}
{\rm Re}\rho^{11}|_{q=0}={\rm Re}\rho^{22}|_{q=0}=&
-\sum_\pm\int \frac{d^3k}{(2\pi)^3}(\pm)\frac{\hbar}{2g_{\rm sd}}[\theta(-\omega_{\bm k\mp})-\theta(-\omega_{\bm k\pm})]
\notag \\
=& \frac{\hbar}{g_{\rm sd}}\frac{k_{F+}^3-k_{F-}^3}{6\pi^2}.
\end{align}
The order of taking the $\bm q,\Omega\rightarrow 0$ limits matters again for ${\rm Re}\rho^{33}$.  This time, $\lim_{\bm q\rightarrow 0}\lim_{\Omega\rightarrow 0}{\rm Re}\rho^{33}\neq 0$ while $\lim_{\Omega\rightarrow 0}\lim_{\bm q\rightarrow 0}{\rm Re}\rho^{33} = 0$.  Again, we determine the order of the limits by the isotropy about $\bm n$.  When ${\rm Re}\rho^{33}|_{q=0}$ does not vanish, $A_t^3 A_t^3$ appears in $\Delta \mathcal S_{\rm eff}$, which breaks the isotropy.   Therefore, contrary to the case of ${\rm Re}\chi^{33}_{ij}|_{q=0}$, we employ $\lim_{\Omega\rightarrow 0}\lim_{\bm q\rightarrow 0}{\rm Re}\rho^{33}=0$ as ${\rm Re}\rho^{33}|_{q=0}$.  Finally, ${\rm Re}\rho^{12}|_{q=0}$ obviously vanishes.  

After all, the expression of the contribution from the conduction electron to spin effective action is 
\begin{align}
\Delta \mathcal S_{\rm eff} =& \int d^4x \Big[
-\hbar \frac{k_{F+}^3-k_{F-}^3}{6\pi^2}A_t^3 + \frac{\hbar^2}{2m}
\left(\frac{\hbar^2}{2m}\frac{1}{\Delta_{\rm sd}}\frac{k_{F+}^5-k_{F-}^5}{15\pi^2}-n\right)[(A_i^1)^2+(A_i^2)^2]
\notag \\
&~~~~~~~~~~~ + \frac{\hbar^2}{2\Delta_{\rm sd}}\frac{k_{F+}^3-k_{F-}^3}{6\pi^2}[(A_t^1)^2+(A_t^2)^2]
\Big]
\notag \\
=&\int \frac{d^3x}{a^3}dt \left[S_c\dot \phi(\cos\theta-1) - \frac{J_cS_c^2}{2}(\partial_i \bm n)^2 + \frac{m_s}{2}\bm{\dot n}^2\right]
\end{align}
where $a$ is the lattice constant and we have defined the spin polarization $S_c$ of the conduction electron per lattice site, the spin-spin exchange coupling $J_c$ and the inertia of spin $m_s$ as 
\begin{equation}
S_c \equiv a^3 \frac{\hbar}{2}\frac{k_{F+}^3-k_{F-}^3}{6\pi^2},~~ 
J_c\equiv \frac{a^3}{S_c^2}\frac{\hbar^2}{4m}\left(n-\frac{\hbar^2}{m}\frac{1}{g_{\rm sd}}\frac{k_{F+}^5-k_{F-}^5}{30\pi^2}\right),~~ 
m_s \equiv \frac{\hbar S_c}{2g_{\rm sd}},
\end{equation}
respectively.  It is easy to see that $J_c>0$ by rewriting it as 
\begin{equation}
J_c = \frac{a^3}{S_c^2}\frac{\hbar^2}{120\pi^2m}\frac{(k_{F+}-k_{F-})^2}{k_{F+}+k_{F-}}(k_{F+}^2+3k_{F+}k_{F-}+k_{F-}^2).
\end{equation}
We may safely neglect the effect of impurity, because the lifetime $\tau$ of the conduction electrons will enter only as $\hbar/(\epsilon_F \tau)$ or $\hbar/(g_{\rm sd} \tau)$, both of which we assume much smaller than one.  

Adding this $\Delta \mathcal S_{\rm eff}[\bm n]$ to the action of localized spin \eqref{spin original action}, we obtain the total effective action $\mathcal S_{\rm eff}[\bm n]$,
\begin{equation}
\mathcal S_{\rm eff}[\bm n] = \int \frac{d^3x}{a^3}dt \left[S\dot \phi(\cos\theta-1) - \frac{JS^2}{2}(\partial_i \bm n)^2 + \frac{m_s}{2}\bm{\dot n}^2\right].
\label{total spin effective action}
\end{equation}
with $S\equiv S_c+S_l$ and $J\equiv(J_cS_c^2+J_lS_l^2)/S^2$.

\section{Derivation of the angular momentum \eqref{angular momentum conservation}}\label{sec: derivation of the angular momentum}

In this appendix, we describe the derivation of the angular momentum current $(\bm j^0, \bm j^i)$ \eqref{angular momentum conservation} from the effective action $\mathcal S_{\rm eff}$ \eqref{spin effective action}, for the readers who are not familiar with such derivation.  Consider a $SO(3)$ rotation in internal space (spin space)\footnote{An element $R$ of $SO(3)$ group and an element $U$ of $SU(2)$ group are related by $R_{ab}\sigma^b=U^\dagger \sigma^a U$ with $\sigma^a$ the Pauli matrices.  We are interested in the classical dynamics of a spin vector, that is, the behavior of $\bm n\equiv \langle \psi |\bm \sigma |\psi \rangle$, where $|\psi\rangle$ is a quantum spin state.  A $SU(2)$ rotation on $|\psi \rangle$ causes a $SO(3)$ rotation on $\bm n$, i.e. $\langle \psi |U^\dagger \sigma^a U|\psi \rangle=R_{ab}n^b$.  }.
  Under this rotation, a vector $\bm v$ in internal space, such as spin $\bm n$ and magnetic field $\bm B$, is rotated\footnote{The magnetic field $\bm B$ forms a scalar product with $\bm n$ in the Zeeman term, which means that $\bm B$ is also a vector in internal space with the same transformation property as $\bm n$, within the framework presented here.}.   When the rotation is infinitesimally small, it is given by $\delta v^a(\bm x,t) = \varepsilon^{abc}\theta^b v^c(\bm x,t)$, where $\theta^a$ are constant infinitesimal transformation parameters.  The effective action \eqref{spin effective action} is invariant under this rotation.  The Noether current corresponding to this invariance, i.e. the angular momentum current, can be obtained by the following trick.  Make the transformation parameters $\theta^a$ be dependent on the position in the real space and time, $\theta^a(\bm x,t)$.  Then, the variation of the action under this space-time dependent internal rotation does not vanish, but is given in the form, 
\begin{equation}
\delta \mathcal S_{\rm eff} = \int \frac{d^3x}{a^3}dt (\partial_0 \bm \theta\cdot \bm j^0 + \partial_i \bm \theta \cdot \bm j^i ) = -\int \frac{d^3x}{a^3}dt~ \bm \theta \cdot(\partial_0 \bm j^0 + \partial_i \bm j^i).
\label{noether}
\end{equation}
Here, terms without any time- and spatial-derivatives on $\theta^a$ are absent, because the rotation would be a symmetry of the action if $\theta^a$ were constants.   We have performed the partial integration in the last equality in eq.\eqref{noether}.  When $\bm n$ satisfies its equation of motion, then $\delta \mathcal S_{\rm eff}=0$ because such $\bm n$ gives a saddle point of $\mathcal S_{\rm eff}$.   Since $\delta \mathcal S_{\rm eff}=0$ for arbitrarily position-dependent $\theta^a(\bm x,t)$, it follows that $\partial_0 \bm j^0 + \partial_i \bm j^i=0$ when $\bm n$ satisfies its equation of motion.  Thus, $(\bm j^0, \bm j^i)$ is the conserved current corresponding to the symmetry under internal rotations, i.e. the angular momentum current.  [In fact, since $\bm n$ is a unit vector, an arbitrary variation of $\bm n$ is given by a rotation in internal space.  Therefore, from the variation \eqref{noether}, the conservation law $\partial_0 \bm j^0 + \partial_i \bm j^i=0$ is the equation of motion itself.]

Concretely, the variation of each term in the effective action \eqref{spin effective action} is given as follows. For the spin Berry phase term, 
\begin{align}
\delta \int \frac{d^3x}{a^3}dt~S\dot \phi(\cos\theta-1) = &
\delta \int \frac{d^3x}{a^3}dt \int_0^1du~ S\bm {\tilde n}\cdot (\partial_u \bm {\tilde n} \times \partial_t \bm {\tilde n})
\notag \\
=& \int \frac{d^3x}{a^3}dt \int_0^1du~S[\partial_u(\bm {\tilde \theta}\cdot \partial_t \bm {\tilde n})-\partial_t(\bm {\tilde \theta}\cdot \partial_u \bm {\tilde n})]
\notag \\
=& \int \frac{d^3x}{a^3}dt~ \bm {\dot \theta}\cdot S\bm n,
\label{variation of the spin Berry phase}
\end{align}
where we have temporarily used $\bm {\tilde n}(\bm x,t,u)$ which is defined over extended space-time $(\bm x,t,u)$ with $0\leq u\leq 1$ a dummy direction and with  boundary conditions $\bm {\tilde n}(\bm x,t,u=0)=\bm n(\bm x,t)$ and $\bm {\tilde n}(\bm x,t,u=1)={\rm const.}$  The transformation parameters $\bm \theta$ are also extended to $\bm {\tilde \theta}(\bm x,t,u)$ similarly. In the last equality in eq.\eqref{variation of the spin Berry phase}, we have performed the partial integration with respect to $t$.  For the inertial term and the exchange coupling term, 
\begin{align}
\delta \int \frac{d^3x}{a^3}dt~ \frac{m_s}{2}\bm {\dot n}^2 
=&\int \frac{d^3x}{a^3}dt~\bm {\dot \theta}\cdot (m_s \bm n \times \bm {\dot n}),
\notag \\
\delta \int \frac{d^3x}{a^3}dt~ \frac{-JS^2}{2}(\partial_i \bm n)^2 
=&\int \frac{d^3x}{a^3}dt~\partial_i \bm\theta \cdot (JS^2 \partial_i\bm n \times \bm n).
\end{align}
The Zeeman term is irrelevant for deriving the angular momentum because it does not involve the derivative on $\bm n$, which means that it does not yield terms proportional to $\partial_0\bm \theta$ or $\partial_i\bm \theta$ in eq.\eqref{noether} under the rotational transformation.  In all, from eq.\eqref{noether}, the angular momentum current can be read as 
\begin{equation}
\bm j^0 = S\bm n + m_s\bm n\times \bm {\dot n}, ~~~\bm j^i=JS^2\partial_i \bm n \times \bm n.
\end{equation}

\section{Derivation of higher order terms in spin effective action}\label{sec: effective action to the fourth}

In this appendix, we derive higher order terms in spin effective action $\Delta \mathcal S_{\rm eff}$ \eqref{Delta Seff to the fourth} in the way discussed in the last part of section \ref{sec: effective action}.  We concentrate only on time-derivative terms, but derivation of higher order terms with spatial-derivatives is essentially the same.   First, all possible terms in spin effective action to the fourth-order are 
\begin{equation}
\Delta \mathcal S_{\rm eff}=\int \frac{d^3x}{a^3}dt \left[S_c \dot \phi(\cos\theta-1) + \frac{m_s}{2}\dot {\bm n}^2 + c_3\bm n\cdot (\bm {\dot n}\times \bm {\ddot n})  + c_4 \bm {\ddot n}^2 + \tilde c_4 (\bm {\dot n}^2)^2\right].
\end{equation}
The angular momentum $\Delta \bm j^0$ can be derived from this action in the way described in Appendix \ref{sec: derivation of the angular momentum}, as 
\begin{equation}
\Delta \bm j^0 = S_c \bm n + m_s \bm n\times \bm {\dot n}  
+ c_3(2\bm{\ddot n}+3\bm{\dot n}^2\bm n) 
+ 2c_4(\bm{\dot n}\times \bm{\ddot n}-\bm n\times \bm{\dddot n}) 
+ 4\tilde c_4 \bm{\dot n}^2 \bm n\times \bm {\dot n}.  
\label{expression of Delta j0}
\end{equation}
This $\bm j^0$ must satisfy $\partial_0\Delta \bm j^0+(2g_{\rm sd}/\hbar)\bm n\times \Delta \bm j^0=0$.  Substitution of eq.\eqref{expression of Delta j0} makes, with $\omega_{\rm sd}\equiv g_{\rm sd}/\hbar$, 
\begin{align}
&\partial_0\Delta \bm j^0+(2g_{\rm sd}/\hbar)\bm n\times \Delta \bm j^0 \notag \\
=&
(S_c-2\omega_{\rm sd}m_s)\bm{\dot n}
+ (m_s+4\omega_{\rm sd}c_3)\bm n\times \bm{\ddot n} \notag \\
&+ (2c_3+4\omega_{\rm sd}c_4)\bm{\dddot n} 
+ (6c_3+12\omega_{\rm sd}c_4)(\bm{\dot n}\cdot \bm{\ddot n})\bm n
+ (3c_3-4\omega_{\rm sd}c_4-8\omega_{\rm sd}\tilde c_4)\bm{\dot n}^2\bm{\dot n}.
\end{align}
The unique solution in order that each coefficient vanishes is that 
\begin{equation}
m_s=\frac{S_c}{2\omega_{\rm sd}}, ~c_3=-\frac{m_s}{4\omega_{\rm sd}}, ~
c_4=-\frac{c_3}{2\omega_{\rm sd}}, ~\tilde c_4 = \frac{5c_3}{8\omega_{\rm sd}}.
\end{equation}

\end{appendix}

\end{document}